\documentclass[fleqn,usenatbib]{mnras}

\usepackage[T1]{fontenc}

\DeclareRobustCommand{\VAN}[3]{#2}
\let\VANthebibliography\thebibliography
\def\thebibliography{\DeclareRobustCommand{\VAN}[3]{##3}\VANthebibliography}

\usepackage{graphicx}	
\usepackage{amsmath}	
\usepackage{multirow}   

\usepackage{newtxtext,newtxmath}

\newcommand{\src}{Cyg~X-1}

\newcommand{\astr}{\emph{AstroSat}}
\newcommand{\nic}{\emph{NICER}}

\newcommand{\rin}{R$_{\rm{in}}$}
\newcommand{\rg}{R$_{\rm{g}}$}
\newcommand{\kte}{kT$_{\rm{e}}$}

\title[Shots in Cygnus X-1 using \astr-\nic\ observation]{Probing the shot behaviour in Cygnus X-1 using simultaneous \astr-\nic\ observation}

\author[Bhargava et al.]{
Yash Bhargava,$^{1}$\thanks{E-mail: yash@iucaa.in}, 
Nandini Hazra$^{2,3,4}$, 
A. R. Rao$^{1,5}$,
Ranjeev Misra$^{1}$, 
\newauthor Dipankar Bhattacharya$^{1}$,
Jayashree Roy$^{1}$ 
and Md. Shah Alam$^{1}$ 
\\
$^{1}$Inter-University Centre for Astronomy and Astrophysics Post Box No. 4, Ganeshkhind, Pune-411007, India\\
$^{2}$Gran Sasso Science Institute, Viale F. Crispi 7, I-67100, L'Aquila (AQ), Italy\\
$^{3}$INFN - Laboratori Nazionali del Gran Sasso, I-67100, L'Aquila (AQ), Italy\\
$^{4}$INAF - Osservatorio Astronomico d'Abruzzo, Via M. Maggini snc, I-64100 Teramo, Italy\\
$^{5}$Department of Astronomy and Astrophysics, Tata Institute of Fundamental Research, Homi Bhabha Road, Mumbai, India
}

\date{Accepted XXX. Received YYY; in original form ZZZ}

\pubyear{2021}

\begin{document}
\label{firstpage}
\pagerange{\pageref{firstpage}--\pageref{lastpage}}
\maketitle

\begin{abstract}

We analyse the aperiodic flaring features, also known as shots, observed in Cyg~X-1 in the 0.1--80~keV energy band using a 6.39 ks simultaneous observation with \astr\ and \nic. We detect 49 simultaneous shots in the soft and hard X-ray bands with \nic\ and \astr-LAXPC, respectively. We observe the shot profile for the first time in soft X-rays (0.1--3~keV), which shows a spectral peak at $\sim$2~keV. Using time-averaged spectroscopy, we measured the truncation of the inner accretion disk at $6.7\pm0.2$ gravitational radii. The shot-phase resolved spectroscopy allowed us to identify the origin of some of the brightest aperiodic peaks in the soft X-rays. We find that the accretion rate is consistent with a constant during the shots while the inner edge of the accretion disk moves inwards/outwards as these shots rise/decay. We discuss the possible mechanisms causing the swing in the inner radius.

\end{abstract}




\section{Introduction}

Persistent Black Hole Binaries (BHBs) are  binary systems comprising a stellar-mass Black Hole (BH) and a high mass companion star feeding it through stellar winds. 
These systems primarily exhibit two states, referred to as the High Soft State (HSS) and the Low Hard State (LHS) \citep{Done2007A&ARv..15....1D}. The Shakura-Sunyaev model of accretion \citep{SS1973A&A....24..337S} provides a good description of the HSS. In HSS, the thermalised accretion disk typically extends to the Innermost Stable Circular Orbit (ISCO). 
The LHS indicates the presence of a truncated disk with a powerlaw component. The powerlaw component is assumed to arise from Compton up-scattering of a fraction of the thermalised disk photons by a hot thermal corona of electrons \citep[and references therein]{Done2007A&ARv..15....1D}. 
The comptonised radiation further interacts with the disk to produce a reflection component, characterised by the fluorescent Fe K$\alpha$ emission and a Compton hump in 10--30~keV \citep{LW1988ApJ...335...57L, Makishima2008PASJ...60..585M, Tomsick2014ApJ...780...78T, Duro2016A&A...589A..14D, Basak2017MNRAS.472.4220B, Tomsick2018ApJ...855....3T}. 

Cygnus X-1 (hereafter \src) is a high mass, persistent X-ray BHB system with an orbital period of about 5.6 days. The bright and persistent nature of the source makes it a promising candidate to study the LHS in BHBs.
Recent measurements indicate that the mass of the BH is 21.2$\pm2.2$~M$_{\odot}$ and the distance to the system is 2.22$^{+0.18}_{-0.17}$~kpc \citep{MJ2021Sci...371.1046M}. The orbital inclination of the binary system is 27.1$\pm0.8^{\circ}$ \citep{Orosz2011ApJ...742...84O} while
\cite{Tomsick2014ApJ...780...78T} found that the inner disk inclination is offset from the binary inclination by $\sim$13$^\circ$ (40.4$\pm0.5^{\circ}$). 
Using the reflection modelling, \citet{Fabian2012MNRAS.424..217F} constrain the spin of the BH to 0.97$^{+0.01}_{-0.02}$, which is also supported by recent  measurements \citep{zhao2020JHEAp..27...53Z, MJ2021Sci...371.1046M}.  The emission from the comptonising medium, seen in the LHS, has been typically modelled 
with a corona either containing two componentsof different optical depths and Compton-y parameters \citep{Makishima2008PASJ...60..585M}  or having  different power law indices and electron temperatures \citep{Basak2017MNRAS.472.4220B, axelsson2018MNRAS.480..751A}. The inner disk radius was measured to be at $\sim$13~gravitational radii (\rg = GM/c$^2$; where M is the mass of the BH) in the LHS of the source by \cite{Basak2017MNRAS.472.4220B} in their best fit model, but the authors concluded that the measurement of the radius depends on the continuum model used to describe the spectrum. \cite{axelsson2018MNRAS.480..751A} have fitted the LHS spectrum with a double comptonisation model and have estimated the inner truncation radius at 7$^{+5}_{-2}$~\rg\ from the time-averaged spectrum and at 5.1$^{+2.3}_{-1.5}$~\rg\ from the frequency-resolved spectrum.  

\src\ is highly variable in the LHS at various timescales.  
Using \emph{Suzaku} observations, \cite{Makishima2008PASJ...60..585M} and \cite{Yamada2013PASJ...65...80Y} studied the spectral properties of the flaring and the non-flaring phases of \src\ in the 0.7--400~keV energy range. The spectrum of the source was modelled with a thermalised disk, the Fe K$\alpha$ emission line (with a gaussian centered at 6.4~keV), a soft and a hard comptonising component using \texttt{compPS} \citep{compPS1996ApJ...470..249P} and a reflection component. 
The studies indicated that the difference between these phases occurs due to either a change in the optical depth of the harder comptonising region, or a change in the temperature of the electrons, or both.

The flaring phases, when observed with high time resolution instruments, have shown a peculiar energy dependence \citep{Negoro1994ApJ...423L.127N, Feng1999ApJ...514..373F, Liu2004ApJ...611.1084L, Yamada2013ApJ...767L..34Y}. 
\cite{Negoro1994ApJ...423L.127N} detected and classified some of the sub-second peaks in the light curve  from \emph{GINGA} observations as shots and studied their averaged profile. The authors selected the peaks which had at least twice the local count rate and were isolated from nearby such peaks by at least 8~s. The averaged profile required two symmetric exponential rise and decay functions for a sufficient description, indicating the presence of two independent timescales. The height of the peak of the co-added shot profile (normalised by the local mean) is observed to decrease with increasing energy. The timescale of the rise of the signal in the  co-added profile is observed to be larger for higher energies, while the decay timescale remains similar across different energy bands \citep{Negoro1994ApJ...423L.127N, Yamada2013ApJ...767L..34Y}, resulting in an asymmetric ratio profile of the shots.  
A phase lag study of the shots by \citet{Negoro2001ApJ...554..528N} associated the longer timescale (0.1--1~s) to the shot variation and the shorter timescale to the spectral hardening. 
Combining the shot detection method of \citet{Negoro1994ApJ...423L.127N} and the $\gamma$-ray burst detection method of \citet{Li1996ApJ...469L.115L}, \citet{Feng1999ApJ...514..373F} detected shots in all the states (LHS, HSS and the transition states between LHS to HSS and HSS to LHS) of \src\ using \emph{RXTE}-PCA observations  and compared the shot parameters across these states. The differences in the timescales of the shots seen in the HSS and the LHS were explained with changes in the size of the comptonising region. \citet{Liu2004ApJ...611.1084L} utilised the algorithm of \citet{Feng1999ApJ...514..373F} on \textit{RXTE}-PCA observations of \src\ in all states to investigate the shot  and shot ratio profiles in HSS, LHS and transition states. \citet{Liu2004ApJ...611.1084L} determined that the hardness ratio profiles in the LHS and transition states follow a trend consistent with the \textit{GINGA} observations by \cite{Negoro1994ApJ...423L.127N}, while the HSS shots showed a turnover in the normalised peak amplitude at $\sim$10~keV. 

Using the high time-resolution instruments onboard \textit{Suzaku} (Hard X-ray detector; HXD and X-ray imaging spectrometer; XIS in p-sum mode), \cite{Yamada2013ApJ...767L..34Y} reproduced the shot analysis done by \citet{Negoro1994ApJ...423L.127N}. The authors determined the peaks greater than the average count rate  in an interval of 1.5~s around the peak and separated from nearby peaks by $>$1.5~s and classified them as shots. The detection criterion used by \cite{Yamada2013ApJ...767L..34Y} is less stringent than the one used by \cite{Negoro1994ApJ...423L.127N} to detect a higher number of shots.   Due to a lack of calibration of p-sum mode data, the authors used the XIS data for detection of the shots and used HXD-PIN and HXD-GSO data for fitting the spectra.  The authors divided the shot profile into 13 non-uniform phase bins and studied the variation of the spectral parameters as a function of the phase. The authors applied the spectral model of \citet{Makishima2008PASJ...60..585M} excluding the soft comptonising component as it was beyond the energy range covered by HXD. 
A shot-phase dependent variation of electron temperature, optical depth and Compton-y parameter was observed, which the authors interpreted as a fluctuation in the comptonising region in the timescale of a shot.

Due to the lack of a time-resolved spectrum in the soft X-rays, \citet{Yamada2013ApJ...767L..34Y} could not model the spectrum with the exact model from \citet{Makishima2008PASJ...60..585M} or \citet{Yamada2013PASJ...65...80Y}, leaving the origin of the shots ambiguous.  
The shot could occur due to a minor increase in the accretion rate, heating up of disk at a timescales of few seconds or due to a random multiplicative process. 
Observations  at a high time resolution in a wide energy band covering the emission from the disk and comptonising medium are required to probe the origin of the shots. 

In this article, we present the analysis of the simultaneous observations by high time resolution instruments \astr-LAXPC and \nic\ covering  the energy range of 0.1--80~keV. The observations and the details of the analysis are described in \S \ref{sec:obs}. The results and a discussion are presented in \S \ref{sec:res}. 

\section{Observation and Data Analysis}\label{sec:obs}

\begin{figure}
\includegraphics[width=\columnwidth]{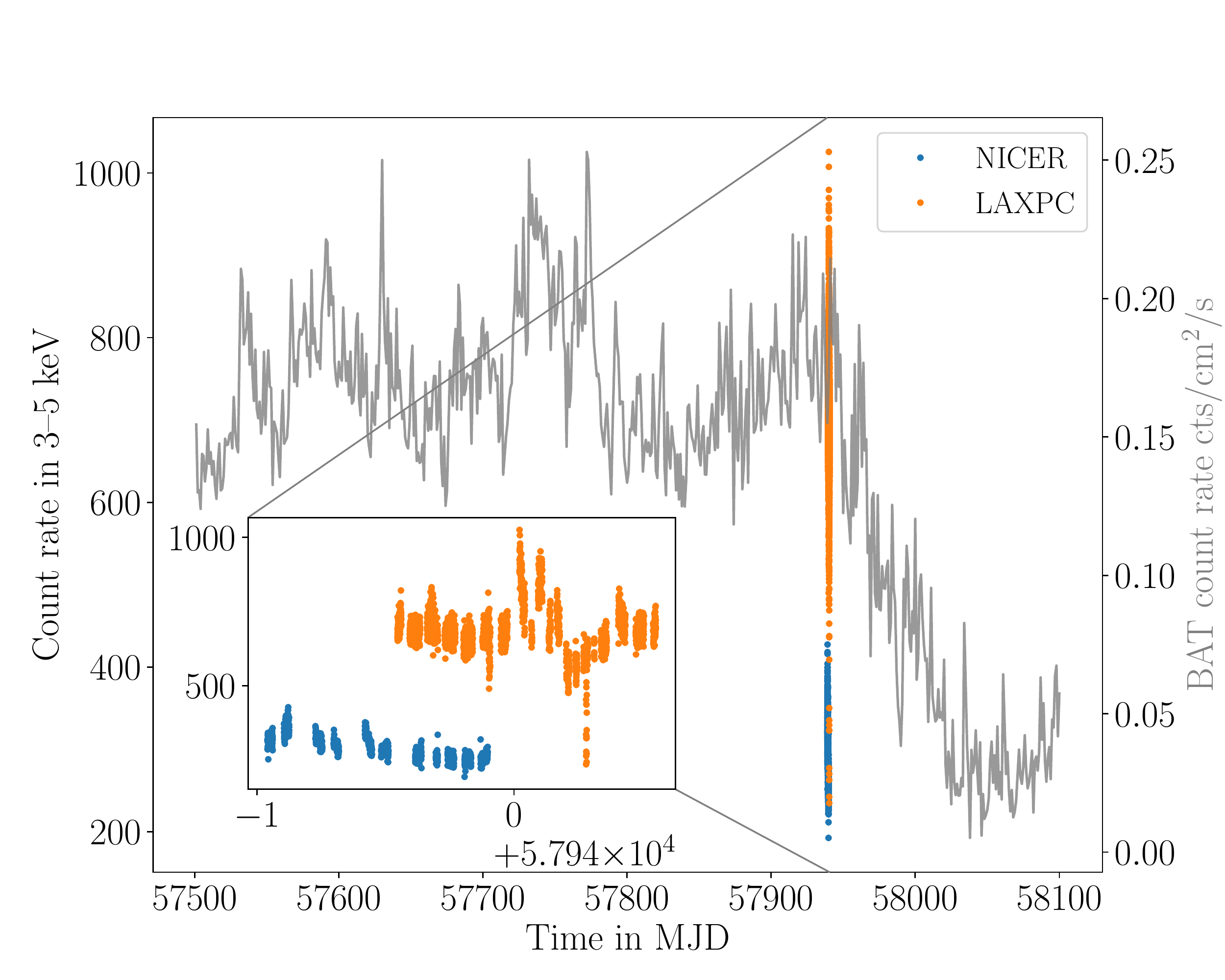}
\caption{The \nic\ and LAXPC observation times are marked as blue and orange points in the Swift-BAT daily light curve of Cygnus X-1, highlighting that the simultaneous observations are made just before a state transition. The  \nic\ and LAXPC  count rates in 3--5~keV and binned at 16~s are shown in the inset (blue and orange points respectively), indicating a significant overlap in the two observations.}

\label{fig:lc_with_bat}
\end{figure}

\src\ was observed simultaneously using \astr\ \citep{Singh2014SPIE.9144E..1SS} and Neutron star Interior Composition ExploreR \citep[\nic,][]{Gendreau2016SPIE.9905E..1HG} from July 4$^{\rm{th}}$ to 5$^{\rm{th}}$ 2017 (see Table \ref{tab:obs} for more details).  We use the data from Large Area X-ray Proportional Counter \citep[LAXPC,][]{Yadav2016SPIE.9905E..1DY, Antia2017ApJS..231...10A} onboard \astr\ due to its large effective area (4000 cm$^2$ at 20~keV for 2 independent active units, LXP10 and 20), high time resolution (10~$\mu$s) and broad energy coverage from 3--80~keV \citep{Antia2017ApJS..231...10A}. \nic\ operates in the 0.1--12~keV energy range and has an effective area of 2000~cm$^2$ at 1.5~keV with a time tagging resolution of $<$300~ns \citep{Gendreau2016SPIE.9905E..1HG}.
To detect and compare features in light curves across two instruments onboard different satellites, we applied barycentric corrections to the timestamps of the individual photons and transformed them to the reference frame  of the  solar barycenter. 
The corrected light curves for both instruments binned at 16~s in the 3--5~keV energy range  are shown in the inset of figure~\ref{fig:lc_with_bat}. The source was observed just before its decline into HSS, as evident from the BAT light curve (plotted in grey). 

\begin{table*}
\centering
\caption{Summary of simultaneous observations analysed in this work. See figure~\ref{fig:lc_with_bat} to see the overlap of the observations.}
\label{tab:obs}
\begin{tabular}{ccccc}
\hline
Observatory-Instrument 	& Observation ID 	& Start MJD 	& End MJD 	& Exposure (Simultaneous) in ks\\ \hline
\hline
\multirow{2}{*}{\astr-LAXPC} &	20170705\_G07\_027T01\_9000001358	&	57939.5465	& 57939.6916		& 10.15 (2.08) \\
		&	20170705\_G07\_042T01\_9000001360	&	57939.6963	& 	57940.5518	& 28.76 (4.31)\\ \hline
\nic\					&	0100320105			&	57939.0408		& 	57939.8974	& 13.21 (6.39) \\
\hline
\end{tabular}
\end{table*}

\subsection{Data reduction} \label{ssec:data_red}

\subsubsection{\astr-LAXPC}

The data from LAXPC observations were reduced using the software \textsc{LAXPCsoftware\_May21}\footnote{\url{http://astrosat-ssc.iucaa.in/uploads/laxpc/LAXPCsoftware_May21.zip}} 
provided by the LAXPC Payload Operation Centre (POC).
We used the tool \texttt{laxpc\_make\_event} to reduce the level~1 data to level~2 data, apply the calibration, and merge different orbits and observations into a single event file. 
We combined the two LAXPC observations into a single event file as \nic\ has simultaneous exposure with both of these and the observations themselves have a slight overlap which is taken into account by the \texttt{laxpc\_make\_event} tool during data reduction. The barycentric correction tool \texttt{as1bary}\footnote{\url{http://www.iucaa.in/~astrosat/as1bary.tar.gz}} was utilised to convert the reference time frame from UTC to Solar system barycenter (TDB). The good time intervals (GTIs), light curves, spectra  and background spectra were extracted by utilising the routines included in the software. The software also determines the appropriate response matrices for  spectral fitting.
The source had enough counts in each bin to use $\chi^2$ statistics but to account for the detector energy resolution ($\sim$10-15~\%), we rebin the spectrum logarithmically with dE/E at 5\% level.

\subsubsection{NICER}

The \nic\ data were downloaded from the HEASARC archive and  were reduced using \texttt{nicerl2} pipeline (distributed with \textsc{HEASOFT} version 6.29c)  utilising the CALDB version xti20210707. The events from the detectors 14 and 34 were removed from the analysis as recommended by the instrument team. The barycentric correction was applied using the tool \texttt{barycorr}. The \textsc{HEASOFT} tool \texttt{xselect} was used to extract the light curves and spectra. The response and ancillary response files were extracted using the \texttt{nicerrmf} and \texttt{nicerarf} respectively.  The background spectrum was computed using the tool \texttt{nicer\_bkg\_estimator} (v0p6) for the common interval, and the same background spectrum was used for each of the shot phases. We compared the background spectrum from an alternative tool \texttt{nibackgen3C50} and  found that the choice of the background model did not affect the results of the spectral fitting due to high count rate of the source. Hence we opted for the former tool. We rebinned the spectrum logarithmically and kept the number of bins similar to that for LAXPC so that the fitting is not dominated by the high number of degrees of freedom from the soft X-ray spectrum and is instead more sensitive to the changes in the high energy tail of the spectrum. This binning method is more conservative than the optimal binning routine suggested by the instrument team.

\subsection{Shot selection}\label{ssec:shot}

We verified that the time series have been correctly aligned after the barycentric correction so that we can detect and compare identical shots across the two instruments. Details of the verification are outlined in Appendix \ref{app: time_match}.

For our study, we redefined the detection procedure of \cite{Yamada2013ApJ...767L..34Y}. The brightness of the source and consequent high count rates in both LAXPC and \nic\ allowed us to use a finer temporal resolution of 0.05~s (compared to 0.1 s in \cite{Yamada2013ApJ...767L..34Y}) to detect the shots. 
Although the temporal resolution of both \nic\ and LAXPC is finer than 0.05~s, we prefer a minimum bin size of 0.05~s as the typical time lag at the Fourier frequencies >1~Hz is less than 0.05~s \citep{Nowak1999ApJ...510..874N, uttley2011MNRAS.414L..60U, demarco2015ApJ...814...50D} across different energy ranges.   
For a simultaneous detection of shots in LAXPC and \nic, we only considered the  common interval across the light curves after barycentric correction. We utilised the light curves covering the complete operational energy range of each instrument for shot detection.
We detected all the peaks in consecutive intervals of $t_{int}$ and determined the highest one. We selected a peak only if the peak count rate was at least $N$ times the local standard deviation above the local mean count rate (both determined from the segment of $t_{int}$) and  separated from the previous selected peak by $M\times t_{int}$ ($M\geq$ 1). We verified our choice of the detection thresholds ($t_{int}$=10~s, $N$=3 and $M$=1) by varying them until the number of peak detections converged. For each peak, we extracted a segment of -10 to 10~s around the peak position and fit the segment with an exponential rise and decay function after subtracting the average of the count rate in the interval -10 to -4 and 4 to 10~s around the peak. 
The rise and decay timescales were allowed to vary independently, and the amplitude of both the exponential functions was tied together. 
To avoid spurious detections, we compared the parameters of each peak as seen by the independent detectors. We used light curves from LXP10 and 20 for parameter comparison. In the case of \nic, we divided the event file into two parts based on the detector ID and compared the light curves from each part. If the parameters of a peak in the two lightcurves  were similar within 20\% of the each other, then we classified the peak as a shot. 
As an additional test of our detection algorithm, we detected shots in LAXPC and \nic\ in 3--5~keV independently and tried various combinations of threshold parameters (viz. $t_{int}$, $N$ and $M$) to determine optimal detection of the shots. We find that similar peaks were being classified as shots but were detected at different significance as the mean count rate was dissimilar owing to different detector sensitivities. For the analysis, however, the full energy range in each instrument was chosen to detect and fit the shots.

In a common exposure of 6.39~ks, we detect 141 shots in LAXPC and 292 shots in \nic\ out of which 49 shots are exactly simultaneous (i.e. the peak time of shot lies within 0.05~s or the bin width of the lightcurve used). The differences in the number of shots detected in the two instruments is due to different operational energy ranges and the effects of  Poisson variation. 

\begin{figure}
    \centering
    \includegraphics[width=\columnwidth]{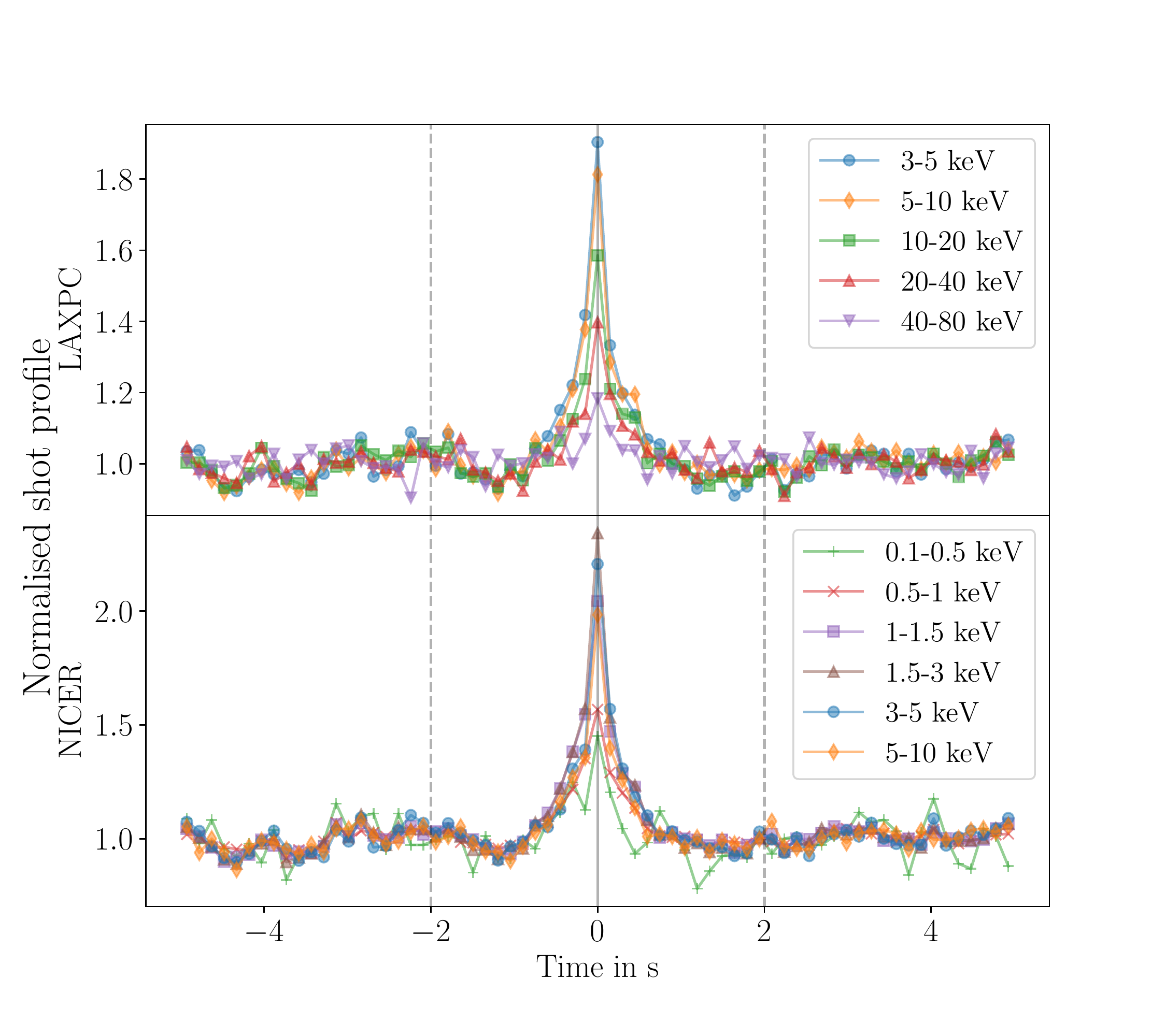} 
    \caption{Co-added shot profiles for LAXPC (top panel) and \nic\ (bottom panel) in different energy bands are shown here. The energy ranges for each instrument are shown in different colours and markers and are indicated in the legend. The shots have been co-added by aligning the peaks and the profile has been normalised by the mean count rate in -4 to -2 and 2 to 4 s interval of the corresponding energy band. The co-adding was done for light curve binned at 0.05~s and was rebinned here (by a factor of 3 in both cases) for clarity. The dashed lines demarcate the interval probed using shot-phase resolved spectroscopy. 	
    }
    \label{fig:shot_rat}
\end{figure}

To demonstrate the spectral dependence of the shots, we extracted the light curves in -4 to 4 s around each shot peak in various energy bands
(3--5, 5--10, 10--20, 20--40 and 40--80~keV for LAXPC; 0.1--0.5, 0.5--1, 1--1.5, 1.5--3, 3--5 and 5--10~keV for \nic). We co-added the shot light curve by aligning the peaks and normalised the co-added peak with the average count rate from -4 to -2 and 2 to 4 s around the peak. The energy bands were picked in order to keep a similar mean count rate across different bands. We plot the shot profiles in figure~\ref{fig:shot_rat}. Noting that the most prominent variation is confined to -2--2~s around the peak, we split the profile in that interval into 9 equal width phase bins and extracted the corresponding GTI. The spectra were extracted for each GTI and fitted as described in section \ref{ssec:shot_spec}.

\subsection{Time-averaged spectroscopy}

To characterise the average behaviour of the source, we modelled the spectrum from 0.7--10~keV for \nic\ and 3.0--50.0~keV for LAXPC, beyond which the background counts were dominating the spectrum. We used a lower limit of 0.7~keV for the \nic\ spectrum as it is dominated by noise and forced trigger peaks in 0.1--0.7~keV band.  
We fit the spectra in \textsc{xspec} (version 12.12.0) using $\chi^2$ statistics. We tied the parameters across \nic\ and LAXPC spectra allowing for a cross normalisation constant. We used an extended energy range of 0.01--1000~keV with 10000 logarithmic bins for accurate computation of the convolution models.

\begin{figure}
    \centering
    \includegraphics[width=\columnwidth]{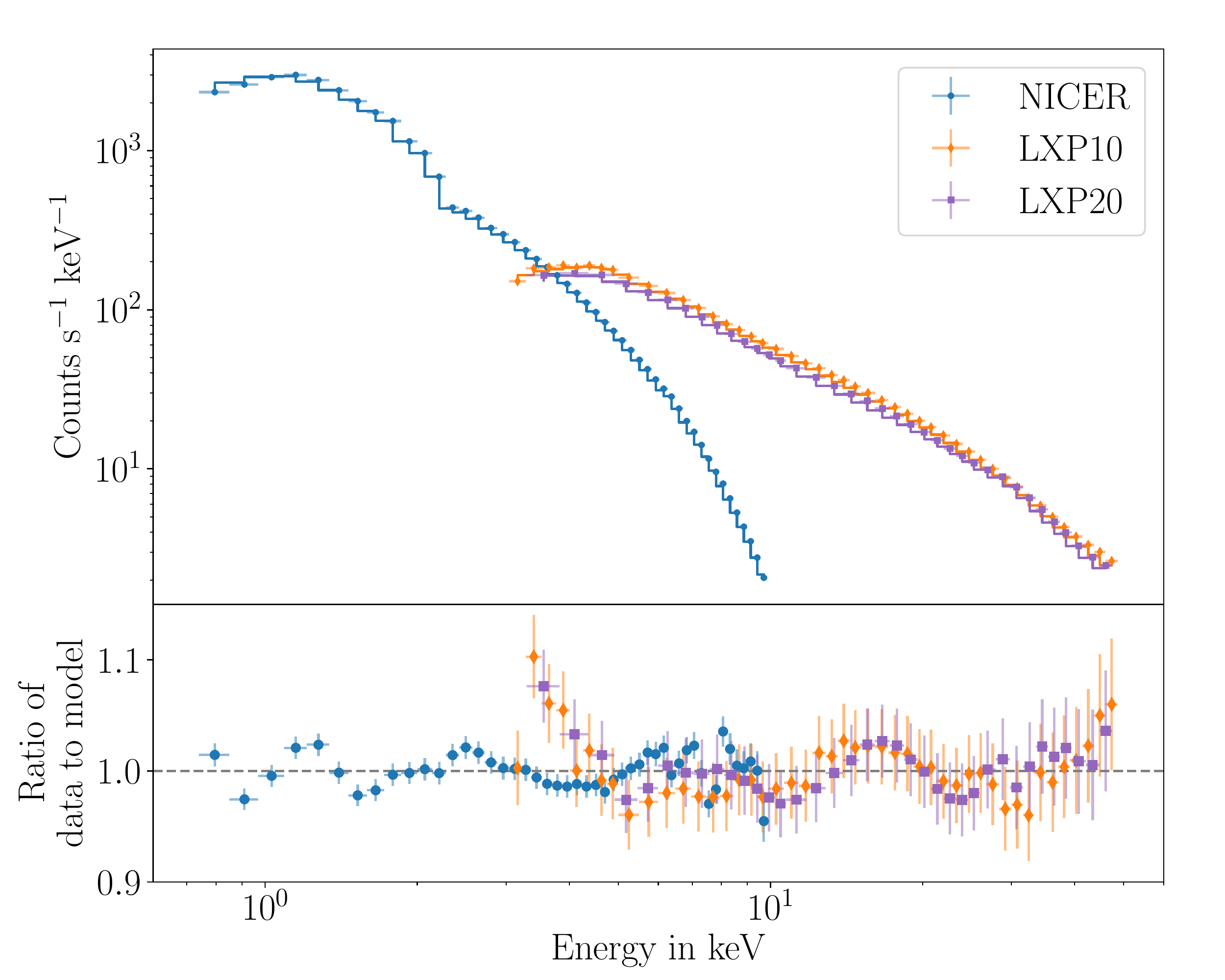}
    \caption{Time averaged spectrum, as observed by \nic\ and \astr-LAXPC, is shown here. The spectra have been logarithmically rebinned and fitted with \texttt{tbabs$\times$(thcomp$\otimes$kerrd + gaussian + ireflect$\otimes($thcomp$\otimes$kerrd-constant$\times$kerrd))}. The fitted parameters are tabulated in Table~\ref{tab:par_vals}. Bottom panel shows the residuals in the form of ratio between the data to best fit model. 
    }
    \label{fig:spec}
\end{figure}

Due to uncertainties in the calibration, we applied additional systematic errors to each spectrum (3\% in LAXPC spectra and 1\% in \nic\ spectra). 
The error bars reported throughout this work correspond to 1$\sigma$ confidence interval unless stated otherwise. 
We assumed that the system has a truncated thermalised disk and a comptonising cloud which intercepts a fraction of the disk photons and up-scatters them to produce the prominent powerlaw tail seen in the spectrum. We also included the interstellar absorption using the model \texttt{tbabs} using the abundances from \cite{Wilms2000ApJ...542..914W} and cross section from \cite{Vern1996ApJ...465..487V}.
The thermalised disk is modelled using \texttt{kerrd} while the thermal comptonisation was modelled using a convolution model \texttt{thcomp} \citep{thcomp2020MNRAS.492.5234Z} using \texttt{kerrd} as an input. As an additional test, we used  \texttt{diskbb} instead of \texttt{kerrd} and found that the fit was identical. However, for interpretation, we used \texttt{kerrd} as it incorporates the spectral hardening factor and gives a more accurate measurement of the inner disk radius. The model \texttt{thcomp} assumes a spherical geometry for the corona. 
We found that the assumed model is  insufficient to explain the spectrum ($\chi^2$ of 561.3 for 123 degrees of freedom, dof) and the residuals indicated the presence of a reflection component. We modelled the reflection using the convolution model \texttt{ireflect}. 
The reflection is assumed to act only on the upscattered photons, while the combination of \texttt{thcomp$\otimes$kerrd} gives both the upscattered photons and the unscattered disk photons. Therefore we include the reflection component by removing the contribution of the unscattered photons from the \texttt{thcomp$\otimes$kerrd} and applying the convolution to the net model. 
On inclusion of the reflection component, the $\chi^2$ reduced to 160.7 for 119 dof. The residuals also indicated a presence of a iron emission line which was modelled phenomenologically with a gaussian line with a fixed width of 0.1~keV. After the inclusion of iron line, the $\chi^2$ reduced to an acceptable value of 132.2 for 117 dof. The best fit model can be written as \texttt{tbabs$\times$(thcomp$\otimes$kerrd + gaussian + ireflect$\otimes($thcomp$\otimes$kerrd-constant$\times$kerrd))}\footnote{Since \textsc{xspec} doesn't support subtraction of models, the constant was actually tied to f-1.} where the constant is tied to 1-f; f being the fraction of photons up-scattered by \texttt{thcomp}. The parameters of all the instances of \texttt{kerrd} and \texttt{thcomp} were tied. The parameters for the fit of the time-averaged spectra are tabulated in Table \ref{tab:par_vals}, while the time-averaged spectrum and the residuals to the best fit model are shown in Figure \ref{fig:spec}.

We fixed the source parameters to the values reported in the literature, i.e. the mass of the BH: 21.2~M$_\odot$, distance to the source: 2.2~kpc \citep{MJ2021Sci...371.1046M}, the inclination of the disk: 40$^\circ$ \citep{Tomsick2014ApJ...780...78T}. We prefer the inclination value from the reflection measurement as compared to the orbital inclination, as it better represents the inner disk inclination.  We fixed the spectral hardening factor of the disk to 1.7 \citep{Shimura1995ApJ...445..780S}. 
We applied a strict lower limit on \rin\ at 1.737~\rg\ as it corresponds to ISCO for a 21.2~M$_\odot$ BH with a spin of 0.97 \citep{MJ2021Sci...371.1046M}. 
We found that the spectrum prefers an electron temperature (\kte) beyond the operational range of LAXPC and since LAXPC has higher systematic uncertainties at higher energies, it cannot constrain the rollover in the spectrum.
We thus froze the \kte\ to 90~keV.  
For this \kte, a corresponding optical depth ($\tau$) can be determined using  equation~14 of \cite{thcomp2020MNRAS.492.5234Z} by switching the sign of the power law index ($\Gamma$) to a negative value\footnote{The $\tau$ is given by the absolute value of the parameter} in the model \texttt{thcomp}.
The measurement of  $\Gamma$ is robust, but the $\tau$ is dependent on the \kte, and thus $\tau$ not be constrained with the current measurement.

\begin{table}
    \centering
    \caption{The best-fit parameter values for fitting of the time-averaged \nic\ and LAXPC spectra to the model \texttt{tbabs$\times$(thcomp$\otimes$kerrd + gaussian + ireflect$\otimes($thcomp$\otimes$kerrd-constant$\times$kerrd))}. The parameters of all instances of thcomp and kerrd are tied.  We use the same model for shot resolved spectra. We find that, across the shot-resolved spectra, certain parameters (indicated below) are consistent with their time-averaged value within 1-2$\sigma$ confidence interval.  We freeze \kte\ at 90~keV as the spectra do not cover the requisite energy range to constrain the parameter. }
    \label{tab:par_vals}
    \begin{tabular}{c|l|r}
    \hline
    Model & Parameter (Unit)                        & Value \\ \hline \hline
    tbabs & N$_{\rm{H}}$ (10$^{22}$ cm$^{-2}$)      &  $0.576^{+0.006}_{-0.005}$* \\ \hline
    \multirow{2}{*}{kerrd} & $\dot{\rm{M}}$  (10$^{16}$ g/s): Accretion rate  & $5.2^{+0.2}_{-0.1} $*     \\ 
          & \rin\        (R$_{\rm{g}}$): Inner radius          & $6.7\pm0.2$        \\ \hline
    
    \multirow{3}{*}{gaussian} & Line Energy     &  $6.41\pm0.03$*\\
            & Line width                  &  $0.1^\dagger$ \\
            & Normalisation            & $(1.9\pm0.3) \times 10^{-3}$* \\ \hline
    \multirow{3}{*}{thcomp} & $\Gamma$: Power law index     &  $1.918\pm0.007$\\
            & f: Scattering fraction                  &  $0.432^{+0.005}_{-0.006}$ \\
            & \kte: Electron temperature  (keV)            & $90^\dagger$ \\ \hline
    \multirow{3}{*}{ireflect} & $\mathcal{R}$: Reflection fraction & $0.72\pm0.04$* \\    
         & $\xi$: Ionisation parameter (erg cm/s) & $167^{+85}_{-55} $* \\ 
         &  Iron Abundance (solar units) & $ 5.5^{+0.8}_{-1.0}$* \\ 
         &  Abundance (solar units) & $ 8\pm2$* \\ \hline 
    \multirow{3}{*}{Flux}  & Disk Flux (ergs cm$^{-2}$ s$^{-1}$) & (1.259$\pm$0.001)$\times$10$^{-8}$ \\
          & Heated Flux (ergs cm$^{-2}$ s$^{-1}$) & (2.191$\pm$0.002)$\times$10$^{-8}$ \\
           & Reflected Flux (ergs cm$^{-2}$ s$^{-1}$) & (6.8$\pm$0.1)$\times$10$^{-9}$ \\ \hline
         & C$_{NICER}$ & 1.21$\pm0.008$ \\
         & C$_{LXP10}$ & 1.02$\pm0.007$ \\
         & $\chi^2/dof$ & 132.2/117\\ \hline
    \end{tabular}
\begin{flushleft}
Note: * The parameter was found consistent with time-averaged value across the shot profile\\
$^\dagger$ The parameter was kept frozen
\end{flushleft}
\end{table}

\subsection{Shot spectroscopy} \label{ssec:shot_spec}

\begin{figure}
    \centering
    \includegraphics[width=0.49\textwidth]{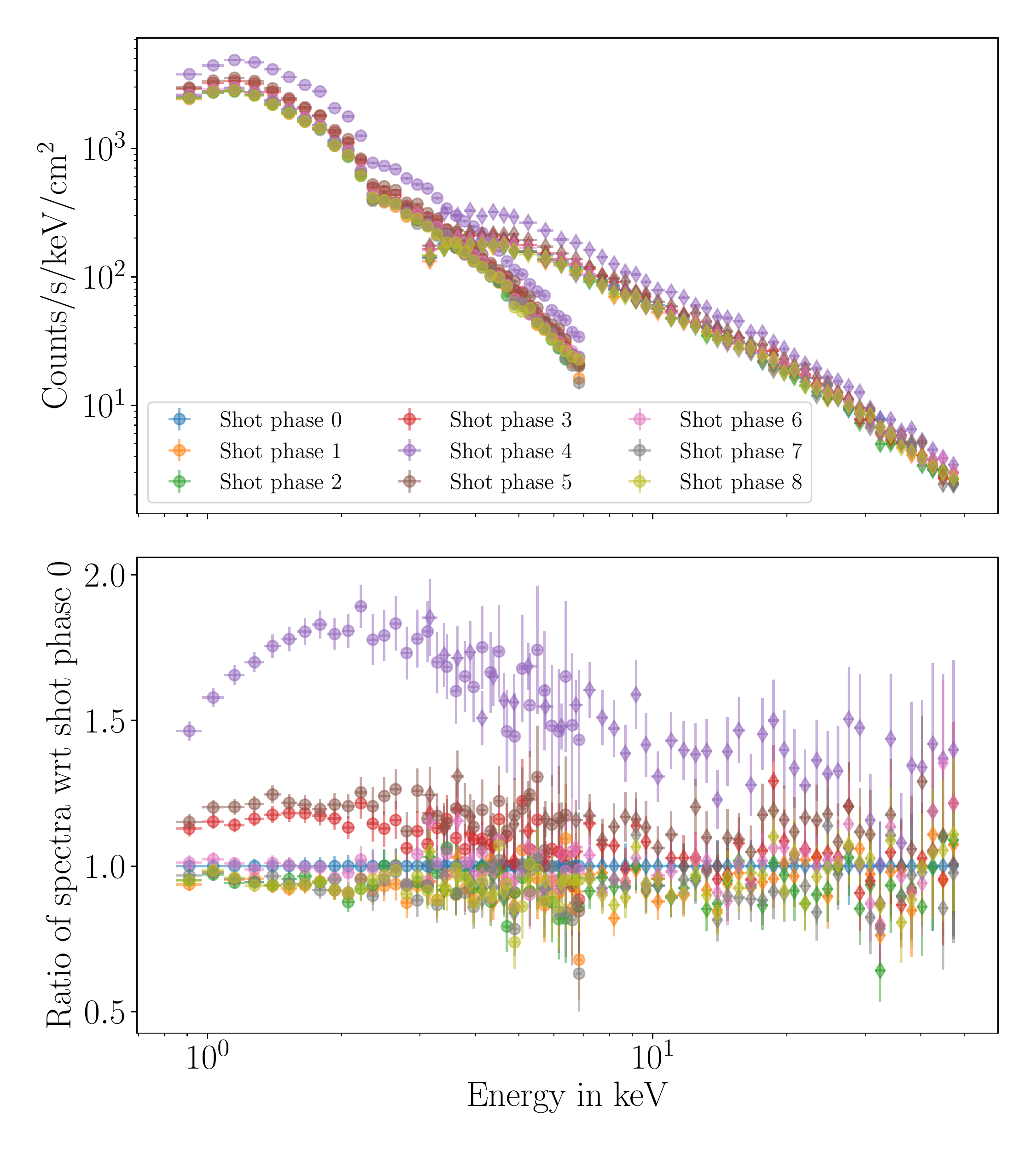}
    \caption{Spectrum for different shot phases as observed by LXP10 and \nic\ are shown in the top panel. To compare the spectra in a model independent way, we also include the ratio of the spectra to the phase-0 spectra in the bottom panel. The spectra for phase 3-5 show clear deviation from 1.0 and spectra for phase 4 show a spectral peak close to 2~keV indicating definite spectral evolution across the shot peak.  }
    \label{fig:spec_rat}
\end{figure}

The variation in the profile is well contained in the interval -2 to 2~s around the peak of the shot.
To study the variation of the spectrum across the shot profile, we divided this interval into 9 uniform phase bins (labelled shot phase 0--8) and extracted the spectrum for each of the phase bins. We used an odd number of phase bins to contain the shot peak in the middle bin and ensure that a slight uncertainty in the determination of the peak does not affect the results. The phase-resolved spectra for the individual shot phases are shown in the top panel of figure \ref{fig:spec_rat}. To highlight the spectral differences in a model independent way, the ratio of the spectra to phase 0 spectra are depicted in the bottom panel of figure \ref{fig:spec_rat}. This also removes the instrumental effects and shows the changes in spectrum across the shot peak.

\begin{figure}
    \centering
    \includegraphics[width=\columnwidth]{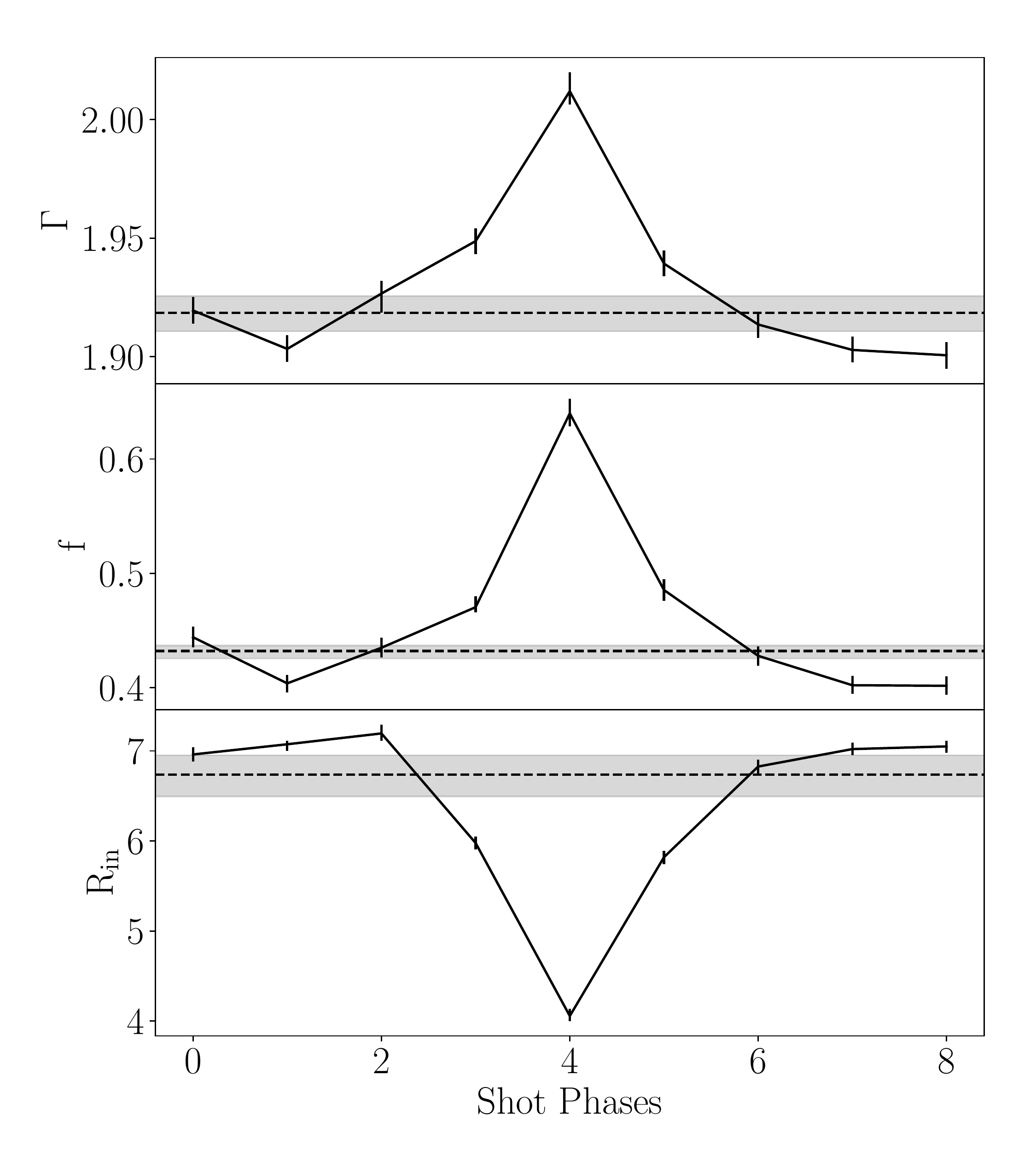}
    \caption{ The variation of parameters ($\Gamma$, f, and \rin\ in top, middle and bottom panels respectively) is shown here. The shot profile corresponds to the interval from -2 to 2~s in  figure~\ref{fig:shot_rat} and is divided into 9 equal phase bins labelled 0--8. 
    The rest of the parameters are found to be consistent with the time-averaged value (within 1$\sigma$) are frozen to that value (see Table \ref{tab:par_vals} for the values of the parameters) and see figure \ref{fig:chi2var} for the variation in the $\chi^2$ as a function of shot phase.  
    The dashed line is the value of the parameter as observed in the time-averaged spectrum, with the shaded region indicating the 1$\sigma$ confidence interval. 
    }
    \label{fig:par_var}
\end{figure}

The spectrum from each phase bin was fitted with the same model as the time-averaged spectrum to study the evolution of the spectral parameters. We froze the iron line parameters for each shot phase to the values suggested by the time-averaged spectrum as the iron line is weakly detected in each shot phase. 
Leaving all the parameters mentioned in table \ref{tab:par_vals} (except \kte) free, we found that many of the parameters are consistent with the time averaged results within 1-2$\sigma$ confidence level. These parameters include reflection fraction ($\mathcal{R}$), Hydrogen column density (N$_{\rm{H}}$), iron abundance, metal abundance and disk ionisation ($\xi$). On freezing these parameters, the $\chi^2$ increases by $\sim$15 for additional 5 degrees of freedom per shot phase spectrum, favouring a simpler model (i.e. the one with fewer parameters free). 
After fixing these parameters, we determined the minimum parameters required to explain the spectral changes across the shot by following the methodology outlined in the appendix~\ref{app:chi2}. 
We find that only $\Gamma$,  f and inner disk radius (\rin) are changing significantly across the shot profile with mass accretion rate ($\dot{\rm{M}}$) being consistent with the time averaged value. 
The variation of $\Gamma$, f and \rin\ after freezing the rest of the parameters to their time-averaged values is shown in figure~\ref{fig:par_var}. The parameter value corresponding to the time-averaged spectrum is indicated as a dashed black line with the shaded region marking the 1$\sigma$ confidence interval reported in Table \ref{tab:par_vals}. Appendix \ref{app:cont} discusses the degeneracy in  determining the value of the key parameters (viz. $\Gamma$, f and \rin) for individual shot phases. 

\begin{figure}
    \centering
    \includegraphics[width=\columnwidth]{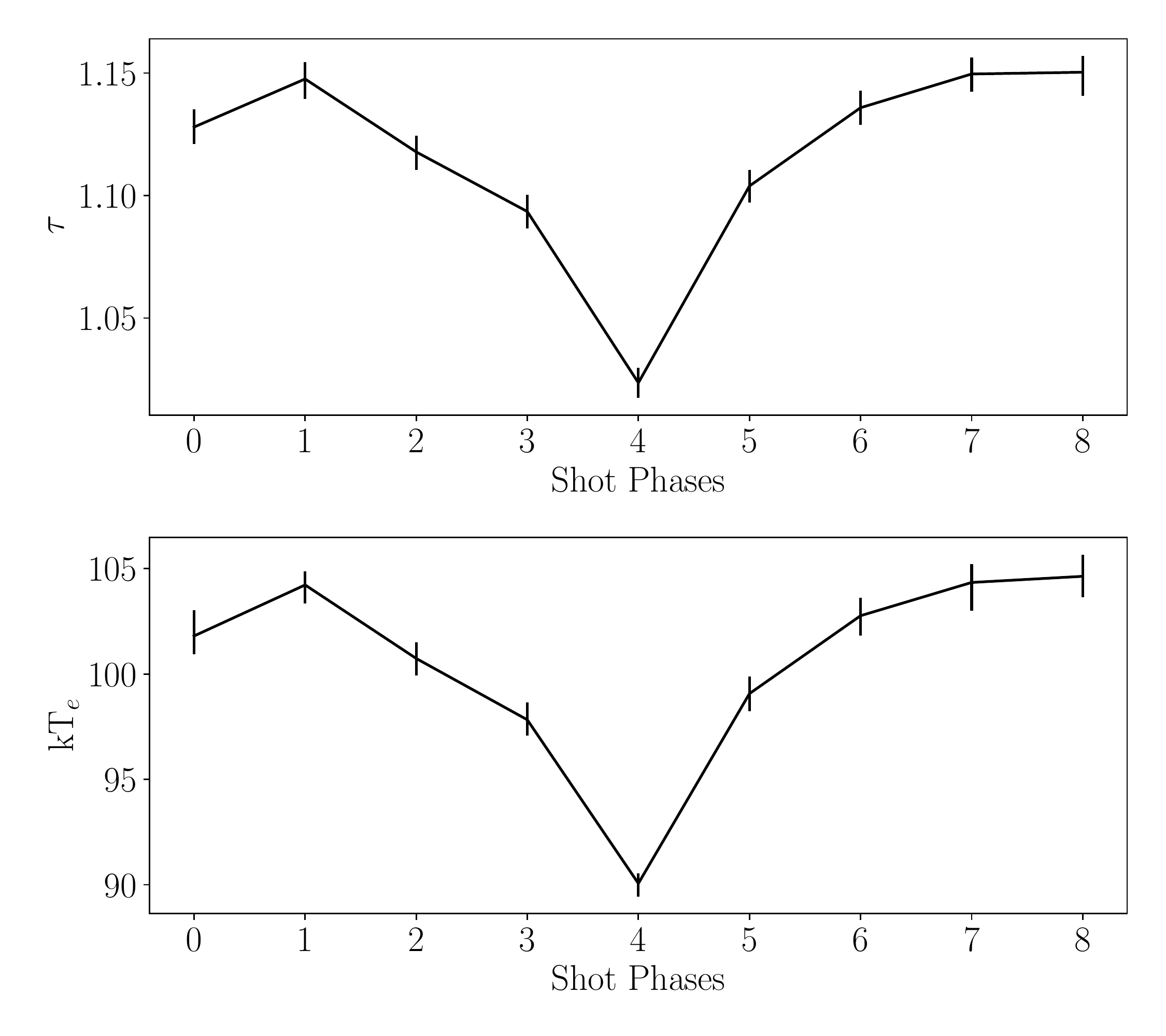} 
    \includegraphics[width=\columnwidth]{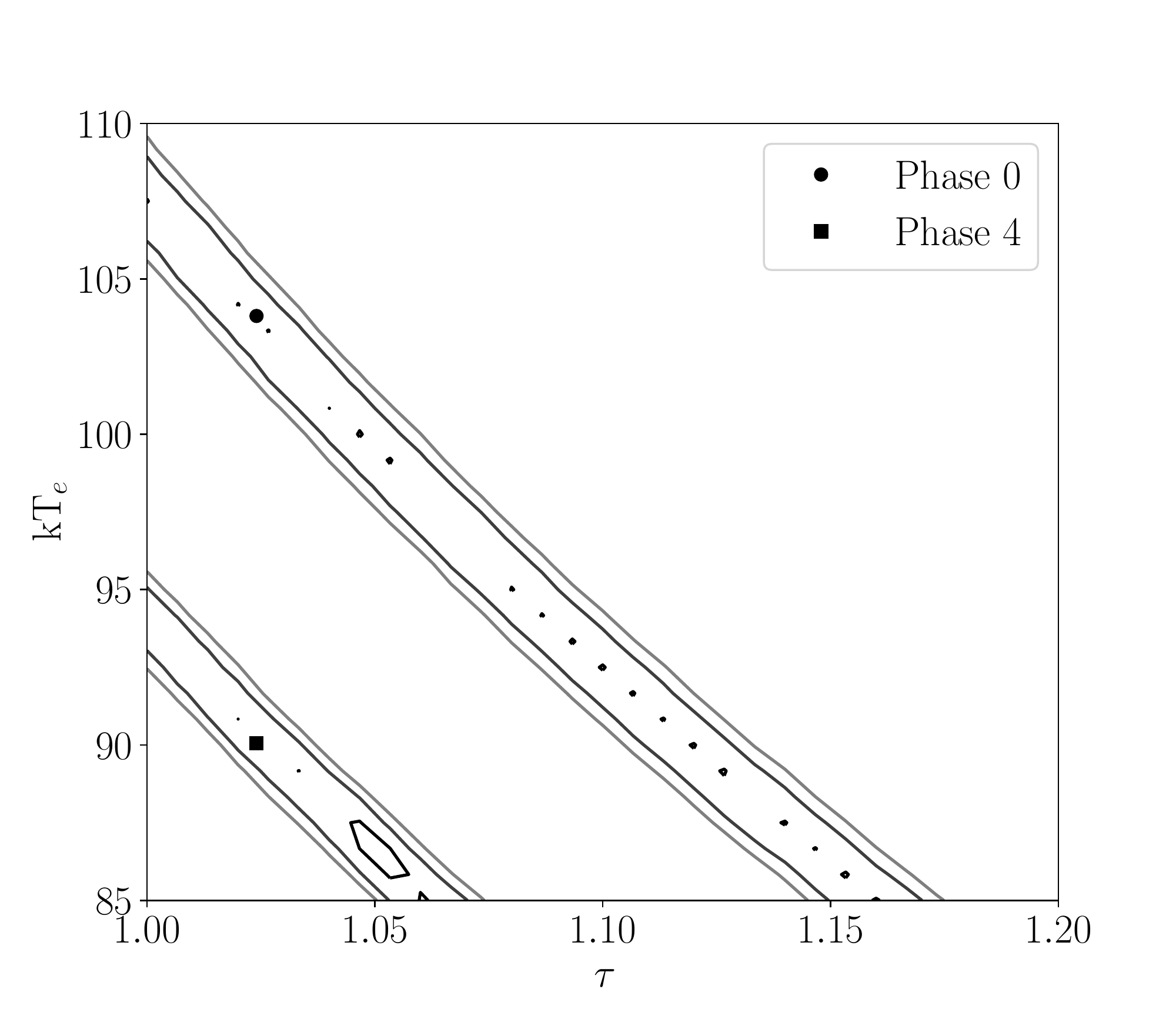}
    \caption{ 
    Testing the degeneracy between the $\tau$ and \kte. In the top panel, the variation of the $\tau$ is allowed to vary across the shot phases while the electron temperature is kept fixed at 90~keV. In the middle panel, the $\tau$ is kept fixed at the value obtained for the peak of the shot (i.e. phase 4) and the electron temperature is allowed to vary. In the bottom panel, the contour plots are shown for simultaneous variation of the $\chi^2$ probing different $\tau$ and \kte. For clarity, the contour plots are shown only for two shot phases 0 and 4. The filled circle and square in the bottom panel indicate the  $\chi^2$ contour corresponding to phase 0 and 4 respectively.   The degeneracy between the parameters is clearly visible which cannot be mitigated without a high energy spectrum to constrain the rollover.  
    }
    \label{fig:tau_test}
\end{figure}

For the observations analysed, we find that the degeneracy between the variation in  \kte\ and  $\tau$ cannot be broken. 
To probe the degeneracy, we determined the variation in $\tau$ and \kte\ independently keeping the other parameter fixed. The variation in $\tau$ was tested by keeping \kte\ fixed at 90~keV (as done in time-averaged analysis). On the other hand, the variation in \kte\ was tested by fixing  $\tau$ for all the phases at the value observed in shot phase 4 (i.e. the peak of the shot) for \kte= 90~keV. In both cases the parameter was observed to reach a minimum at the shot peak and the value of the parameter dropped by $\sim$15\% at the shot peak. We also computed the $\chi^2$ by varying $\tau$ and \kte\ in a grid slightly larger than the variation observed in the top panel of figure~\ref{fig:tau_test}. The contours of the $\chi^2$ variation for two shot phases (0 and 4) are shown in bottom panel of the figure~\ref{fig:tau_test}.  

\begin{figure}
    \centering
    \includegraphics[width=\columnwidth]{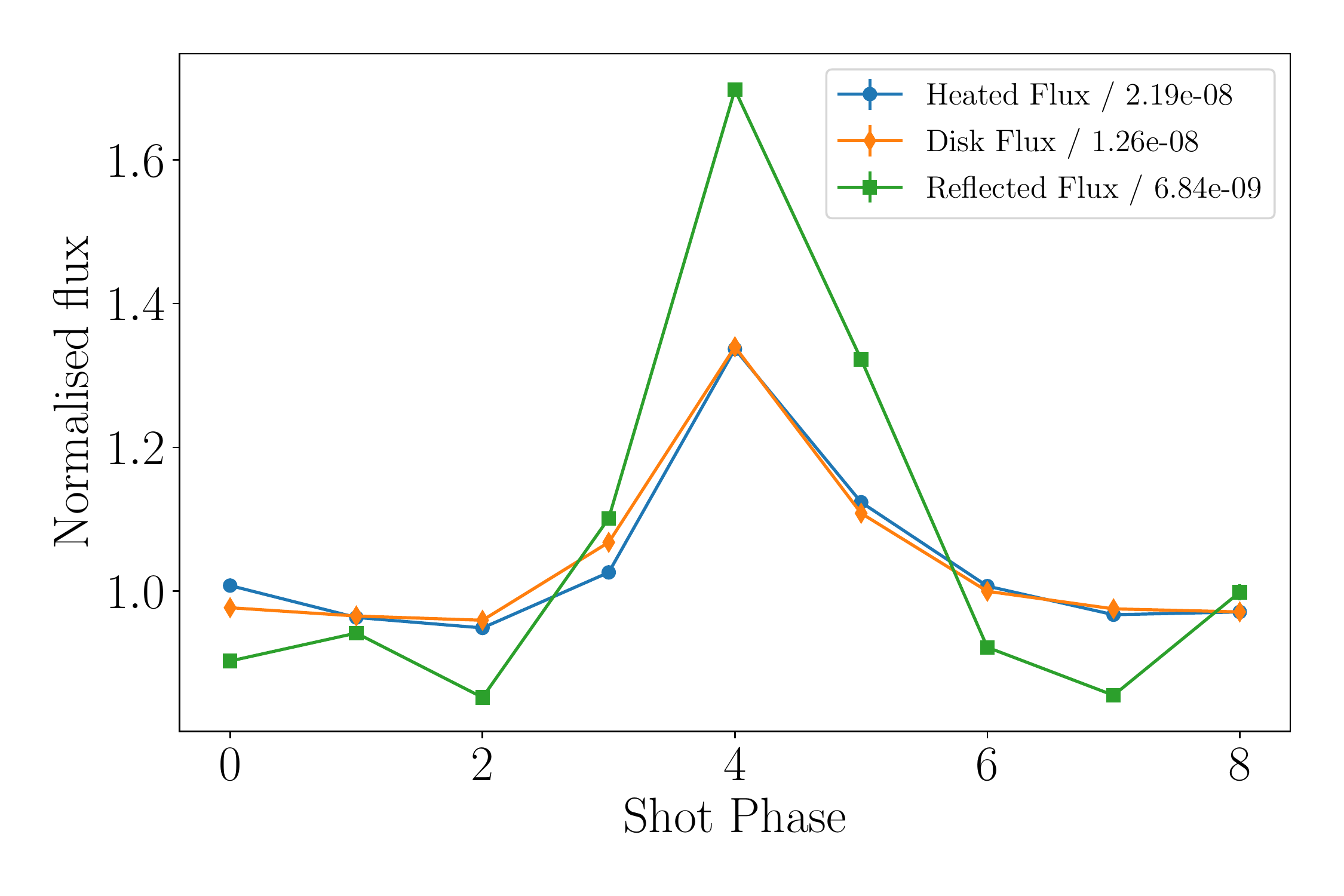}
   
    \caption{ The variation of the flux of different spectral components across the shot profile is shown here. To highlight the relative variation, the fluxes have been normalised by the value mentioned in the legend.  The heated flux is obtained from difference of \texttt{cflux} on \texttt{thcomp$\otimes$(gaussian+kerrd)} and \texttt{cflux} on \texttt{kerrd}. The disk flux is obtained from \texttt{cflux} on \texttt{kerrd}. The iron line component is  constant across the shot profile and the flux contribution from is about 2 orders lower than the measured disk flux. The heated flux is heating provided by the comptonising medium and is computed by subtracting the previous fluxes. 
    The fluxes also follow the show profile indicating that during the shot, both fluxes from the disk and the heating provided by the comptonising cloud rises.}
    \label{fig:fluxes}
\end{figure}

We computed the fluxes from the disk and the comptonised region using \texttt{cflux} model in \textsc{xspec}. The fluxes were determined in 0.1--500~keV band by applying \texttt{cflux} on \texttt{kerrd} and \texttt{thcomp$\otimes$(kerrd)}. The difference in the two fluxes corresponds to the amount of heat supplied by the comptonising medium. We also computed the flux from the reflection component by applying  \texttt{cflux} on the \texttt{ireflect$\otimes($thcomp$\otimes$kerrd-constant*kerrd)}. The variation of the  fluxes as a function of the shot phases is shown in  figure~\ref{fig:fluxes}. The fluxes are normalised to the corresponding value obtained from the time-averaged spectrum to highlight the variation better.   

\section{Results and Discussion} \label{sec:res}

Using simultaneous observation of \astr\ and \nic, we study the spectral properties  of \src. The variability in \src\ is known to span decades in frequency \citep[see][for example]{Maqbool2019MNRAS.486.2964M} and is coupled across different frequency regimes \citep[][]{Uttley2005MNRAS.359..345U}. In this work, we devise a peak isolation technique based on the procedure outlined in \citet{Yamada2013ApJ...767L..34Y} to identify and characterise some of the variability in \src.  In a common exposure of 6.39~ks, we detect 49 simultaneous peaks in LAXPC and \nic\ which we characterise as shots by comparing the properties of the peak across independent lightcurves and across different instruments. Compared to \citet{Yamada2013ApJ...767L..34Y}, we detect far fewer shots. We attribute this to our strict constraint of simultaneous detection of peaks in LXP10, LXP20 and \nic\ with near-identical shape (quantified by the fitting of each peak and comparing the parameters across the independent detectors). These shots represent the brightest and most prominent peaks in the lightcurve detected in a wide energy range (see Appendix \ref{app:ccf} for a comparison between cross-correlation functions from shot intervals and complete lightcurve). 
The variation of the shot profile across different energies as seen by LAXPC is quite similar to that observed by \citet{Yamada2013ApJ...767L..34Y} and \citet{Liu2004ApJ...611.1084L}. These authors have also observed a decrease in the relative shot peak amplitude with energy and thus a sharper dip in the ratio profile.
The  profile of the shots in soft X-rays, observed for the first time with \nic, paints a slightly different picture. The relative peak height is the smallest at the lowest energy bin 0.1--1~keV, rises till 2~keV and then drops again (see the ratio plot in figure \ref{fig:spec_rat}).  This is similar to the peaking of the relative shot profile (at $\sim$10~keV) as observed by \citet{Liu2004ApJ...611.1084L} for the HSS shot analysis. The timescales seen in the \nic\ profile follow similar trends as seen in the LAXPC data. The differences in the relative amplitude and in the ratio of the shot resolved spectra point to  spectral changes in the source in the timescale of the shot. Due to a fewer number of shots detected, we are unable to reproduce the asymmetric shot ratio profile observed in \cite{Yamada2013ApJ...767L..34Y}.  

\subsection{Time averaged results}
We model the time-averaged spectra using a thermalised accretion disk with a comptonising cloud of hot electrons up-scattering a fraction of the disk photons and the emission from the comptonising cloud then reflected from the cool disk. Additionally, we find evidence of a broad iron line, which has been reported previously \citep{Makishima2008PASJ...60..585M, Duro2016A&A...589A..14D, walton2016ApJ...826...87W}. We find that the absorption column density of neutral hydrogen is $\sim$5.7$\times10^{21}$~cm$^{-2}$. \citet{Parker2015ApJ...808....9P}, \citet{walton2016ApJ...826...87W} and \citet{Basak2017MNRAS.472.4220B} have observed an N$_{\rm H}$ of 4--7$\times10^{21}$~cm$^{-2}$ for different epochs of the source indicating a variable nature of the absorption column due to the stellar winds in the source.  
We observe a truncated disk ($\sim$6.7~\rg) with a $\dot{\rm{M}}$ of 5.2$\times10^{16}$~g/s. The \rin\ measurement is larger than the one seen in \citet{Parker2015ApJ...808....9P} but smaller than that of \cite{Basak2017MNRAS.472.4220B}. An important caveat regarding the measurement of the inner disk radius is that it depends significantly on the assumed mass of the BH. The measurements by \cite{Parker2015ApJ...808....9P}, \cite{Duro2016A&A...589A..14D}, \cite{Basak2017MNRAS.472.4220B} and \citet{axelsson2018MNRAS.480..751A} have assumed 14.8~M$_\odot$ BH \citep{Orosz2011ApJ...742...84O} which would lead to a higher estimate of inner truncation radius. For example, the inner radius computed from the \texttt{diskbb} normalisation from \citet{Duro2016A&A...589A..14D} is $\sim$99~km (or $\sim$3.17~\rg; after incorporating the spectral hardening factor of 1.7). For a BH of 21.2~M$_\odot$ at a distance of 2.2~kpc \citep{MJ2021Sci...371.1046M} the same measurement  becomes consistent with our result. Furthermore, \citet{Basak2017MNRAS.472.4220B} emphasise that the measurement of the inner disk radius depends strongly on the continuum assumed. 

The powerlaw tail is modelled with thermal comptonisation by a spherical corona with a high electron temperature. We find that typically 40\% of the photons from the disk are intercepted by the comptonising medium. \citet{Parker2015ApJ...808....9P} find that the spectral cutoff in the \textit{Suzaku} spectrum is observed at about 100~keV. This is beyond the energy range of LAXPC, and  due to  reduced sensitivity at higher energies, we are unable to constrain the electron temperature. We find that for a fiducial value of 90~keV, we get an acceptable fit to the spectrum.   
We obtain a significant reflection fraction ($0.43^{+0.005}_{-0.006}$), which is consistent with \citet{Makishima2008PASJ...60..585M}. Different studies have shown different measurements of the reflection fraction \citep{Tomsick2014ApJ...780...78T, Parker2015ApJ...808....9P, Duro2016A&A...589A..14D, Basak2017MNRAS.472.4220B}, which could be attributed to a source evolution or differences in the assumptions of the models or differences in the interpretation of the reflection fraction \citep[e.g. ][]{dauser2016A&A...590A..76D}. 
Additionally, \citet{Parker2015ApJ...808....9P} note a degeneracy between the reflection fraction and the iron abundance.  
We observe that the disk is weakly ionised similar to \citet{Makishima2008PASJ...60..585M} but in contrast to the high ionisation observed in \citet{Tomsick2014ApJ...780...78T}, \citet{Parker2015ApJ...808....9P}, \citet{Duro2016A&A...589A..14D}, \citet{walton2016ApJ...826...87W} and \citet{Basak2017MNRAS.472.4220B}. 
We observe  supersolar iron abundance in the source, which has also been previously reported \citep[][]{Parker2015ApJ...808....9P, Duro2016A&A...589A..14D, Basak2017MNRAS.472.4220B}. 
While \cite{Makishima2008PASJ...60..585M} and \cite{Basak2017MNRAS.472.4220B} both require two-component comptonisation models to fit the high-resolution spectrum from \emph{Suzaku} and \emph{Suzaku}-\emph{NuSTAR} observations, we find that an adequate fit can be obtained with a single comptonising zone. 

\subsection{Variation of parameters across the shot}
Across the shot, we observe that N$_{\rm{H}}$, $\dot{M}$, abundances, $\xi$, and $\mathcal{R}$ are consistent with the values observed in the time-averaged spectra. 
The spectral parameters $\Gamma$,  f and \rin\ of the phase bins 0--2 and 6--8 are close to the time-averaged parameters indicating that the processes corresponding to the shot are limited to phase bins 3--5 (i.e. -0.833--0.833~s around the shot peak). Within the reduced interval,  $\Gamma$ and f rise as the shot peak approaches and decay after the peak. The inner disk radius, however, reduces as the shot peak approaches
and then rises back to the time-averaged value as the shot decays. Across the shot, the accretion rate of the system is consistent within 4\% of the measured value.
\citet{Yamada2013ApJ...767L..34Y} have described the shot phenomenon in 10--200~keV energy range, which is primarily dominated by the emission from the comptonising region and the reflection. Thus assuming a near symmetric inflow of soft photons, the authors have described that due to the asymmetric response of the comptonising region (i.e. electron temperature and optical depth), the hardness ratio of the shot is asymmetric. In our work, we have compared the shots in the complementary energy range of 0.1--80~keV and observed that the shot profiles are indeed symmetric. Due to a lack of sensitivity and energy coverage at hard X-rays, we are unable to observe the asymmetric ratio profiles and the observed variation in both electron temperature and optical depth. Combining the results of \citet{Yamada2013ApJ...767L..34Y} and present work, we describe the shot phenomenon as follows. At a constant accretion rate, the disk moves suddenly inwards, heating up in the process. As the temperature rises, the emission from the disk increases (as evident from the variation of the flux, Figure~\ref{fig:fluxes}). The surge of photons then interacts with the comptonising medium causing a change in the powerlaw index, which is a manifestation of a simultaneous change in optical depth and electron temperature as shown by \citet{Yamada2013ApJ...767L..34Y}. We also note that during the shot, the fraction of photons intercepted by the comptonising medium also increases and then decreases, possibly related to the change in the inner radius of the disk. The flux contributions of the disk and the comptonising medium are observed to mimic the shot profile.

For an accretion disk,  the thermal timescale at a particular radius can be estimated by calculating the disk parameters from the observed physical parameters. A rough estimate indicates that the thermal timescale at 6.7~\rg\ is about 0.14~s, which is faster than the timescales in which shot occurs. This indicates that our assumption of the thermalised disk is valid even for each shot phase. 

The inward and outward motion of the disk is observed to happen within a few seconds, with the accretion rate remaining  constant within 4\%. The inward/outward motion of the disk is often assumed to be the difference between different states of \src\ \citep{Done2007A&ARv..15....1D} and is usually linked with the changes in the accretion rate. \citet{meyer2007A&A...463....1M} and \citet{taam2018ApJ...860..166T} have described the phenomenon with condensation/evaporation of the matter in the comptonising medium onto the accretion disk. Although the processes described in these works are used to model the transition between different spectral states of the source, a temporary condensation/evaporation could also explain the motion of the inner disk without a change in the accretion rate. Detailed calculations regarding this mechanism is beyond the scope of this work.

\section{Conclusions}

We probe the peaking features observed in \src\ in 0.1--80~keV energy range using high time resolution instruments LAXPC and \nic. Our results can be summarised as follows:
\begin{enumerate}
    \item Observation of the shot profile in soft energy bands for the first time: The shot profile in different energy bands in soft X-rays was observed for the first time using the \nic\ observation. The shot profile indicates that the variation of the relative amplitude of the shot peak is not monotonic, and the amplitude is maximum at $\sim$2~keV energy bands. 
    \item Measurement of inner radius: Assuming a thermalised disk with a spherical corona up-scattering a fraction of photons and reflection of the comptonised photons from the disk, we measure the inner truncation radius at 6.7$\pm0.2$~\rg. 
    \item Origin of the photon surge: Using the wideband simultaneous coverage, we fit the shot-phase resolved spectra and determine that the sudden spike in the count rate is related to the inward motion of the inner edge of the disk, which leads to a larger fraction of photons being scattered and a subsequent cooling of the electron cloud. 
   \end{enumerate}

\section*{Data availability}

The \nic\ data is available at \url{https://heasarc.gsfc.nasa.gov/docs/archive.html}. The \astr\ data is available at \url{https://astrobrowse.issdc.gov.in/astro_archive/archive/Home.jsp}. In both cases, the datasets can be searched in the respective websites using the observation IDs listed in Table \ref{tab:obs}.

\section*{Acknowledgements}
This work makes use of data from the \astr\ mission of the Indian Space Research Organisation (ISRO), archived at Indian Space Science Data Centre (ISSDC). The authors would like to acknowledge the support from the LAXPC Payload Operation Center (POC) TIFR, Mumbai  in data reduction. YB would like to thank Diego Altamirano, Liang Zhang and \nic\ operations team for their help in understanding  the \nic\ tools. YB would also like to thank Tomaso Belloni for the help in understanding the GHATS software and the aperiodic behaviour in the lightcurves of BHBs. 
This research has made use of data and/or software provided by the  High  Energy  Astrophysics  Science  Archive  Research  Center (HEASARC), which is a service of the Astrophysics Science Division at NASA/GSFC and the High Energy Astrophysics Division of the Smithsonian Astrophysical Observatory.




\bibliographystyle{mnras}
\bibliography{ref} 



\appendix

\section{Matching LAXPC and \nic\ times}\label{app: time_match}

\begin{figure}
    \centering
    \includegraphics[width=0.8\columnwidth]{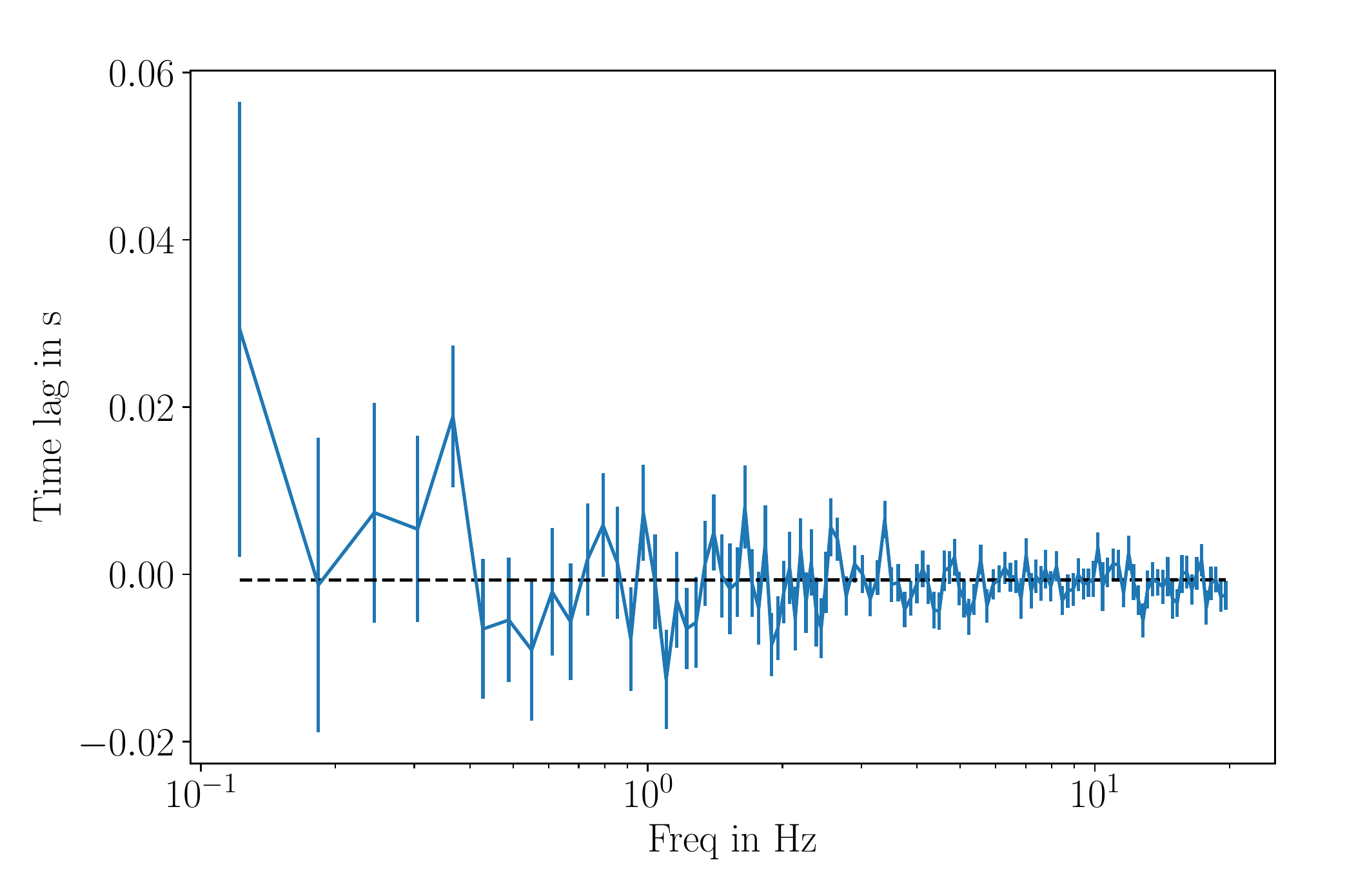}
    \includegraphics[width=0.8\columnwidth]{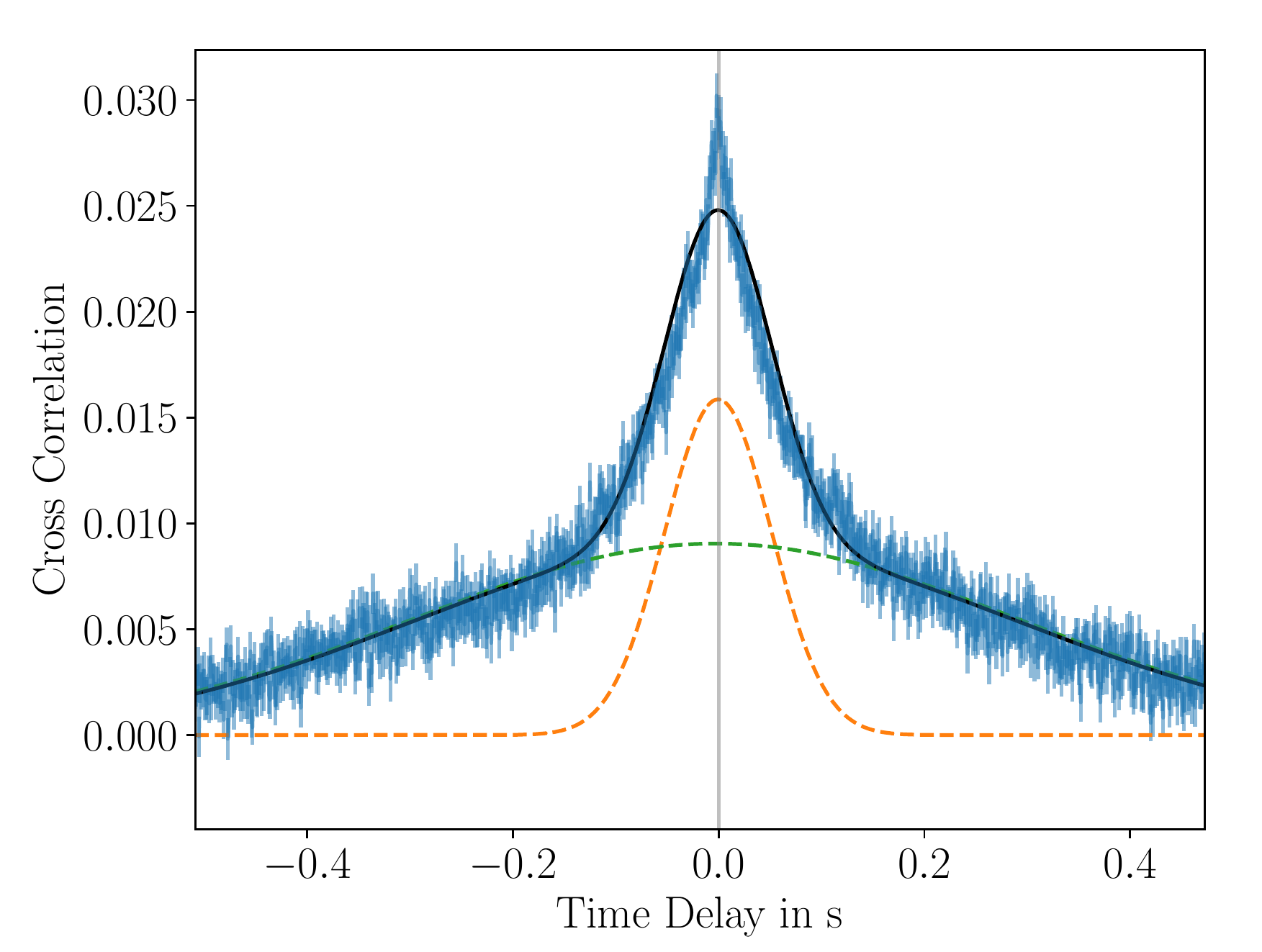}
    \caption{Cross correlation (top panel) and Frequency dependent time lag (bottom panel) between \nic\ and LAXPC 4--8~keV. The cross correlation function has been fitted with two Gaussian functions of different width (shown as orange and green dashed lines). The combined model is shown as black curve. The peaks of both functions are consistent with 0.0 within 2$\sigma$. The frequency dependent time lags were fitted with a constant which is also consistent with 0.0 within 2$\sigma$. }
    \label{fig:ni_lxp_lag}
\end{figure}

To have identical detection of shots across two different instruments it is paramount to verify if the variability across the detectors is identical. To verify if we have successfully aligned the time series, we computed the cross-correlation between the light curves of the same binning from LAXPC and \nic\ in the same energy range (4--8~keV). We ensured that only the exact simultaneous light curves are being cross-correlated by creating standard Good Time Intervals (GTIs) using the tools provided by the instrument team (\texttt{laxpc\_make\_stdgti} for LAXPC and \texttt{nimaketime} for \nic). The standard GTI were then merged using the ftool \texttt{ftmgtime} with "AND" option. The cross correlation for light curves of different time resolutions is found to peak at zero lag. The peak of the correlation was fitted with two Gaussians. The peak of both Gaussians are consistent with zero within 2$\sigma$ ($(-3.2\pm 2.02) \times 10^{-3}$~s and $(-4.93\pm 4.93) \times10^{-4}$~s for orange and green curves in figure~\ref{fig:ni_lxp_lag} top panel respectively). 

We also computed the coherence as a function of Fourier frequencies and computed the phase lags \citep{Nowak1999ApJ...510..874N}. For a identical detection the phase lags should be consistent with zero. The time lag in the frequency range 0.1--20 Hz is consistent with zero within 2$\sigma$ ((-7.2$\pm4.6$)$\times10^{-4}$~s, see figure~\ref{fig:ni_lxp_lag} bottom panel). The value of the lag is much smaller than minimum time bin used in the analysis and thus we assume that the observations have been successfully aligned. 

\section{Variability of the shot intervals vs variability in the full lightcurve} \label{app:ccf}
\begin{figure}
    \centering
    \includegraphics[width=0.8\columnwidth]{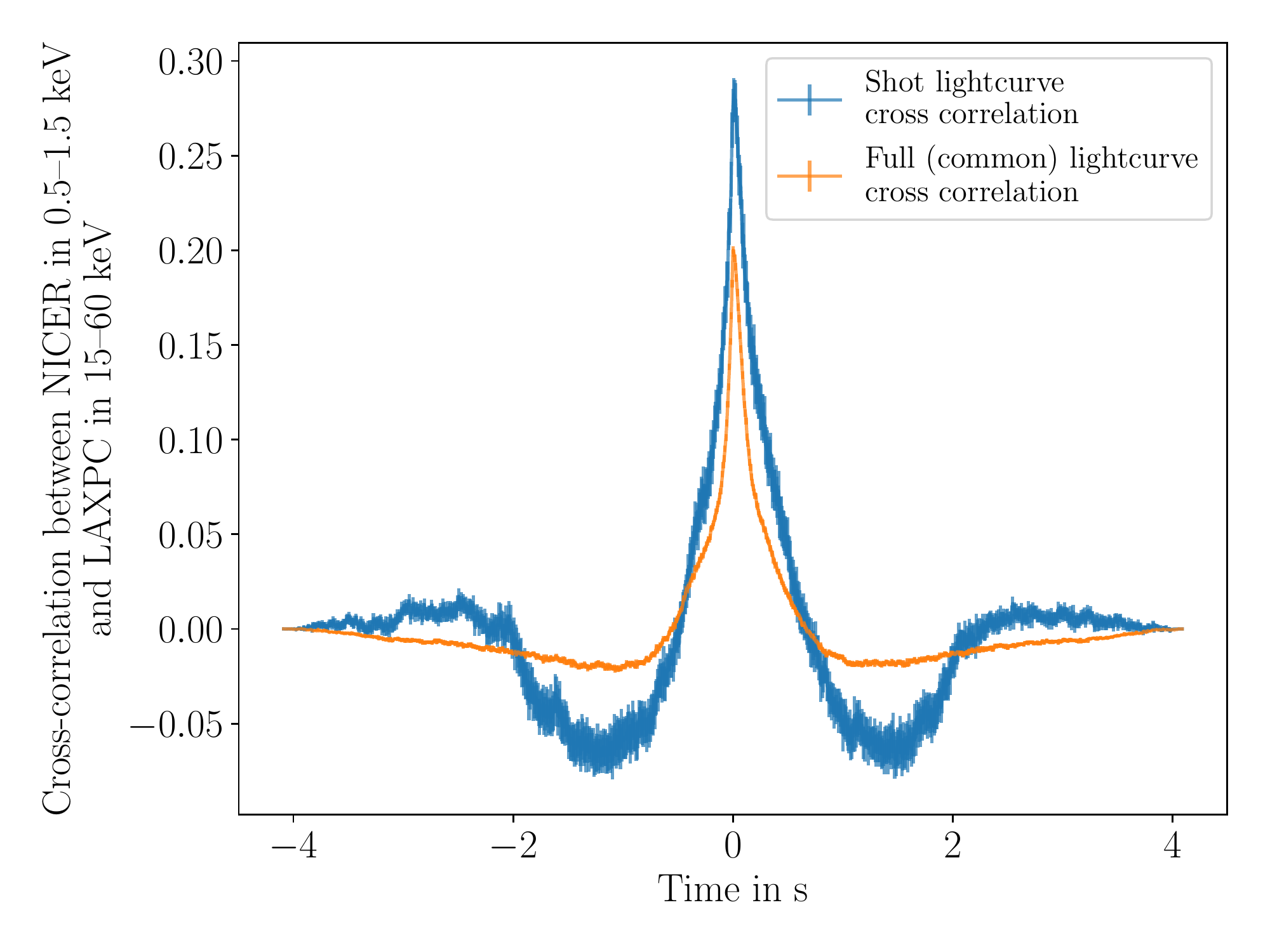}
    \caption{Cross-correlation function (CCF) between NICER lightcurve in 0.5--1.5 keV and LAXPC lightcurve in 15--60 keV. The CCF for the shot lightcurves is shown in blue while the CCF for the common lightcurve is shown in orange. The CCFs differ significantly for different time delays but peak at 0.0~s.  }
    \label{fig:ccf}
\end{figure}

To determine if the intrinsic variability during the shot interval is significantly different from the general aperiodic variability, we computed cross-correlation  functions during shot intervals and the full lightcurve between extreme energy bands (0.5--1.5 keV for NICER and 15--60~keV for LAXPC, see figure \ref{fig:ccf}). We find that the CCFs are significantly different indicating that the variability probed by the brightest and most significant shots represents a part of the general aperiodic variability.  

\section{Determining minimum number of parameters for shot spectroscopy} \label{app:chi2}

\begin{figure*}
    \centering
    \includegraphics[width=0.48\textwidth]{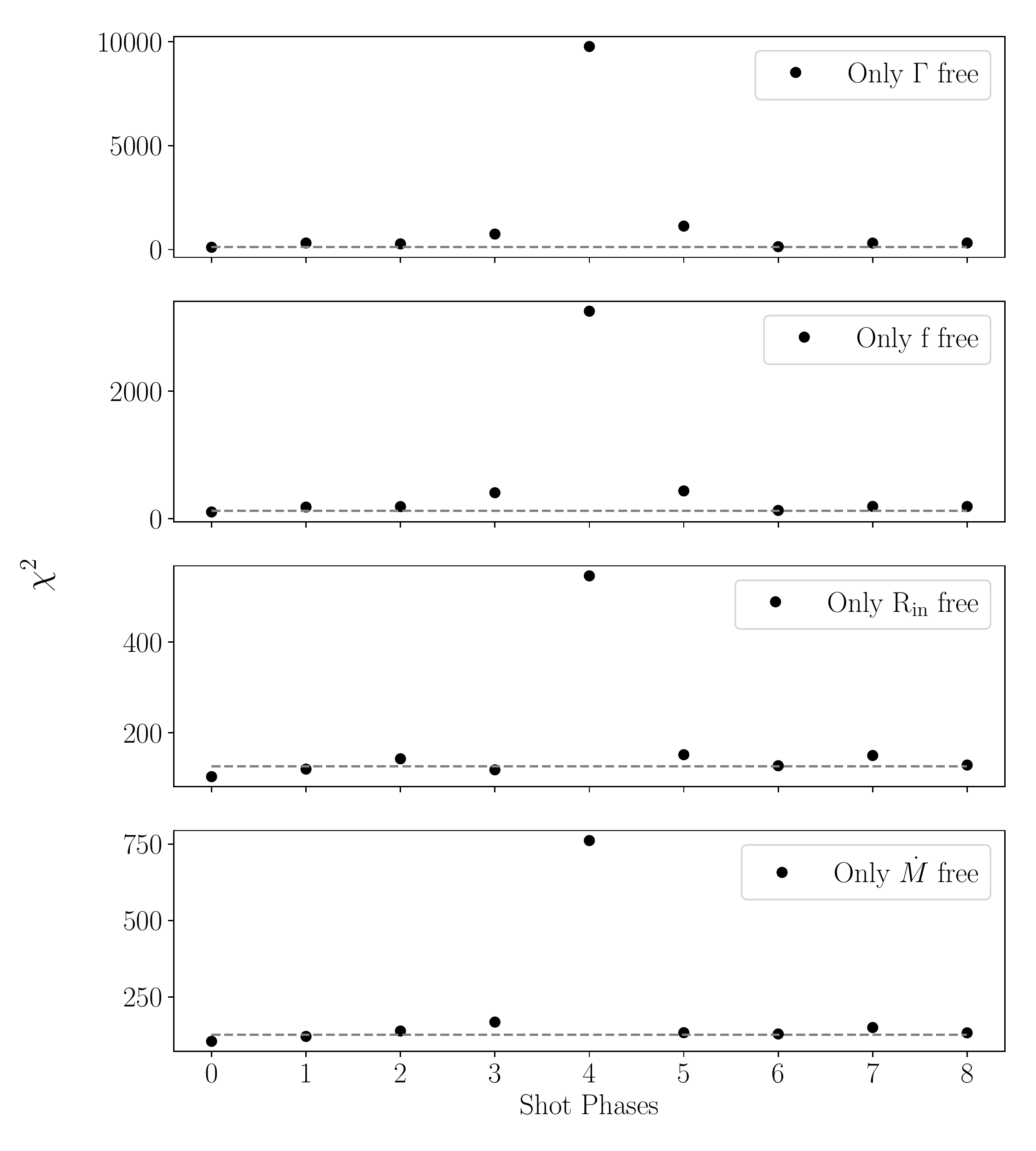} \includegraphics[width=0.48\textwidth]{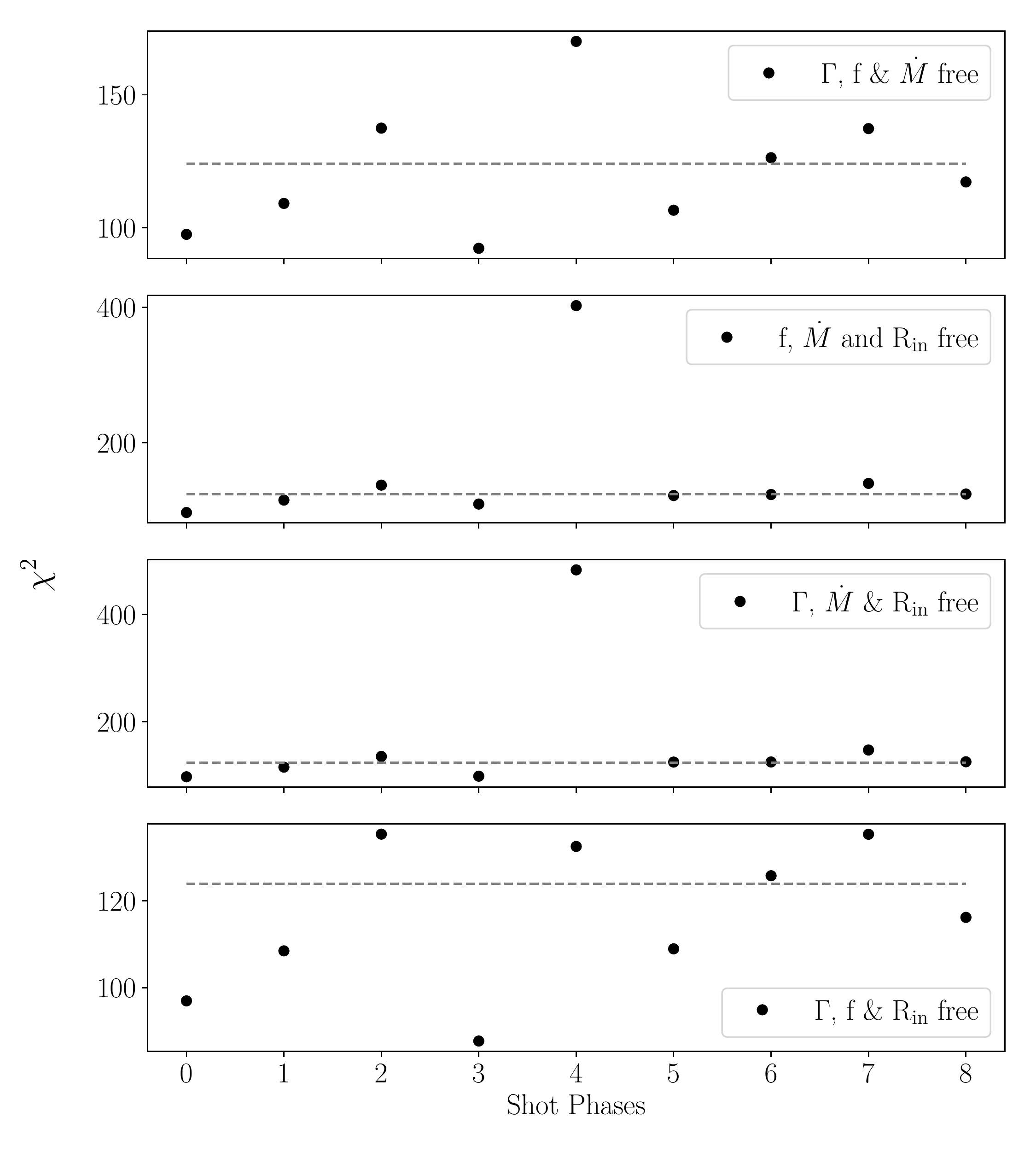}
    \includegraphics[width=0.8\textwidth]{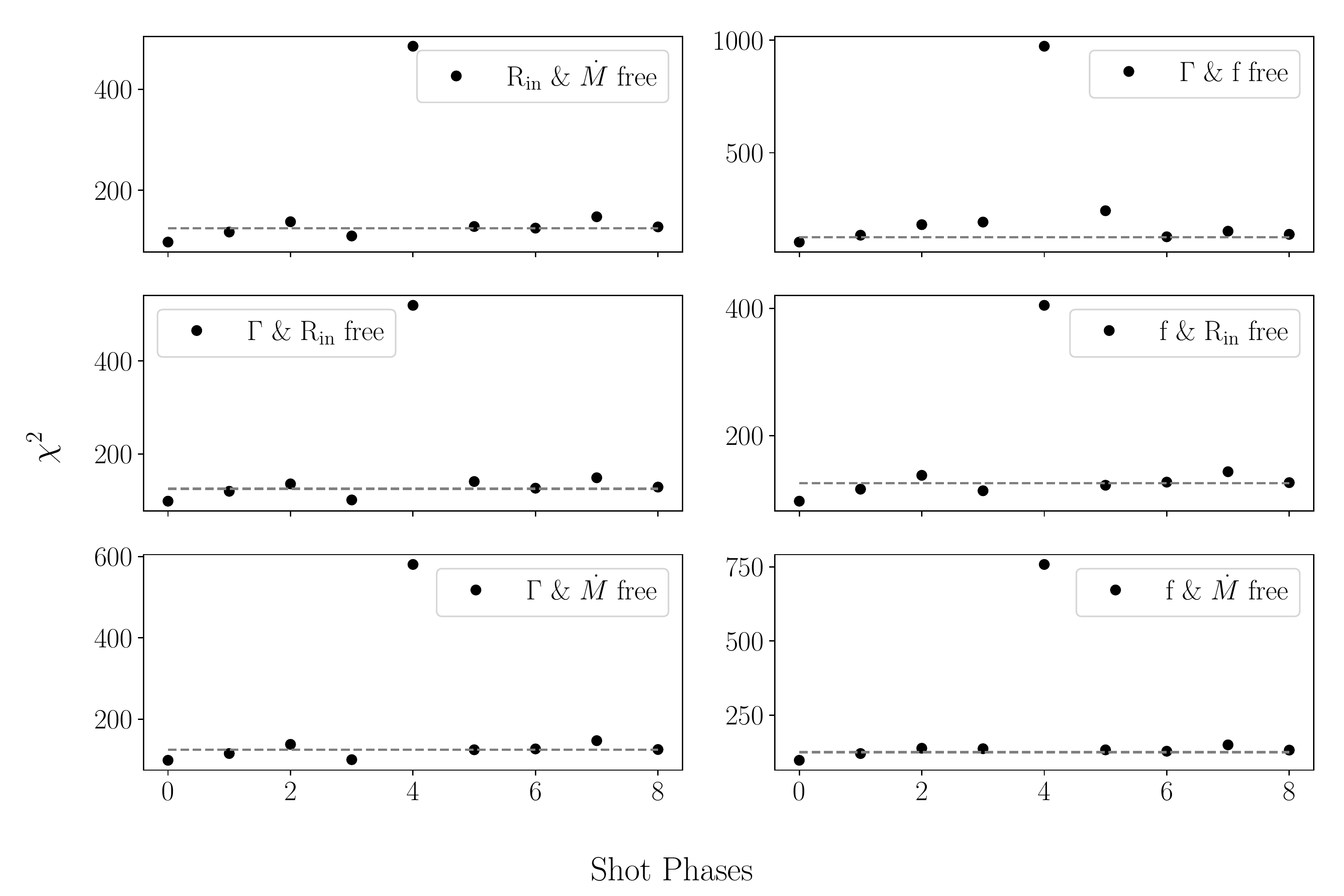}
    
    \caption{Variation of Best-fit $\chi^2$ as a function of shot phase while keeping various combinations of parameters free. The $\chi^2$ values are shown as black dots while the degrees of freedom are shown as a grey dashed line. The parameter(s) kept free are indicated in the legend of respective panels.}
    \label{fig:chi2var}
\end{figure*}

To determine the minimum number of parameters which could explain the spectral changes during the shot we observed the variation in the $\chi^2$ by varying different combinations of the $\Gamma$, f, \rin\ and $\dot{\rm{M}}$ and keeping the rest fixed to the value suggested in the time averaged fits (table \ref{tab:par_vals}). 
The variations of the $\chi^2$ and the corresponding degrees of freedom are shown in figure \ref{fig:chi2var}. 
A high $\chi^2$ is indicative of a bad fit and therefore assumed model being insufficient to explain the variation. In all cases (except one) tested like this, we find that the $\chi^2$ is high at the shot peak and in single parameter cases, it is unacceptable for shot phases 3 and 5. The only case where the $\chi^2$ is acceptable for all cases, is when $\Gamma$,  scattering fraction (f) and inner disk radius (\rin) are left free and ($\dot{\rm{M}}$) is kept fixed at time-averaged value. 

\section{Probing degeneracy between $\Gamma$, f and \rin} \label{app:cont}

\begin{figure}
    \centering
    \includegraphics[width=\columnwidth]{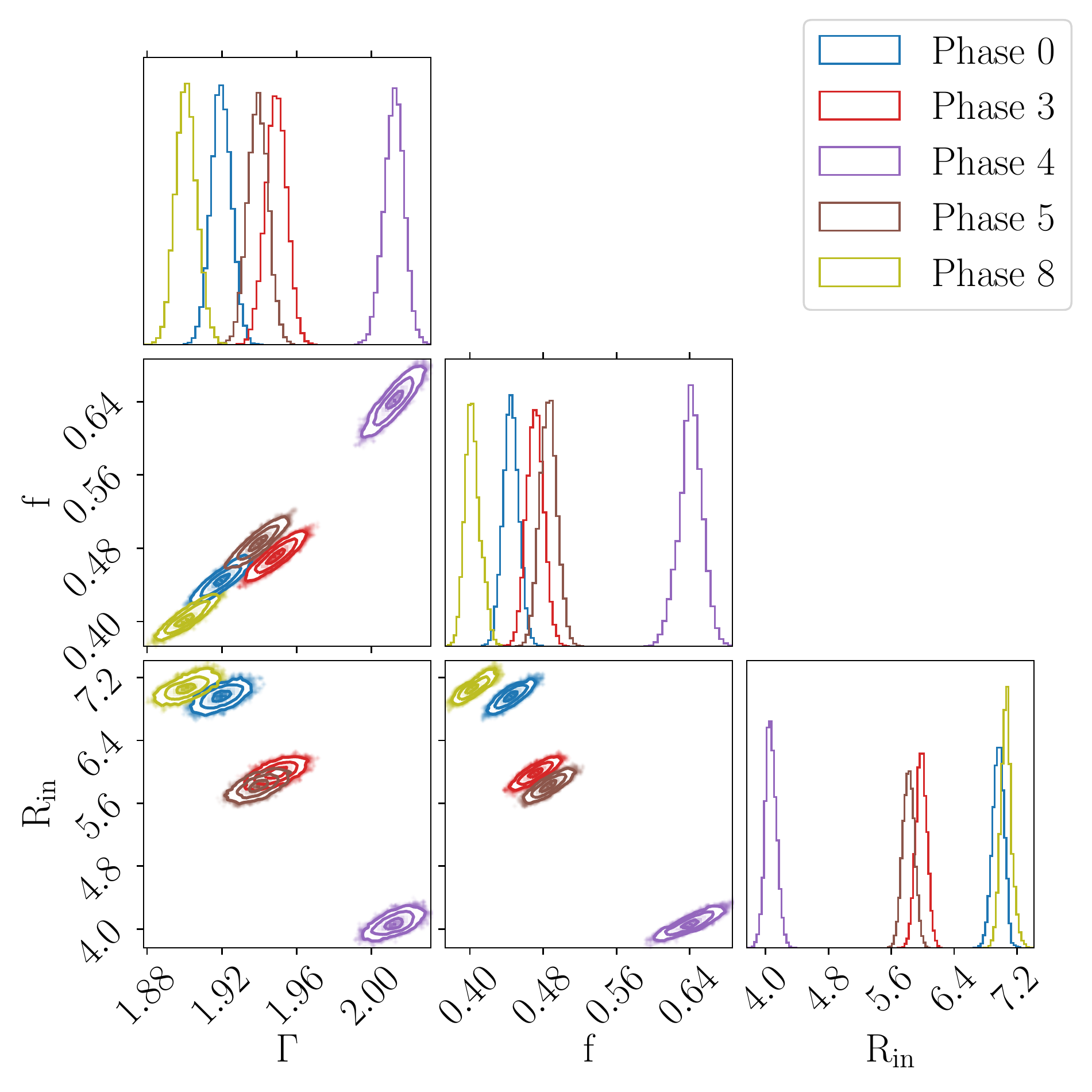}
    \caption{1D and 2D distribution of the parameters for different shot phases. The distribution is shown only for five shot phases for clarity. The colour scheme is kept consistent with figure \ref{fig:spec_rat}. }
    \label{fig:contours}
\end{figure}

For each shot-phase spectrum, we sampled the $\chi^2$ function by allowing $\Gamma$, f and \rin\ to vary. The parameter space was sampled using a Markov Chain Monte Carlo sampler with Goodman-Weare algorithm. We initialised 50 walkers for 1000 steps each (after removing 100 steps corresponding to the burn-in period). The walkers were initialised using the parameter values and the covariance matrix obtained after fitting the spectra in \textsc{xspec}.  The 2D and 1D distribution of the parameters are shown as a corner plot \citep{corner} for some of the shot phases in figure \ref{fig:contours}. 

\bsp	
\label{lastpage}
\end{document}